\documentclass[iop]{emulateapj}
\usepackage{subfigure}

\shorttitle{M31 RR Lyr}
\shortauthors{Jeffery, et al.}

\begin{document}

\title{HST/ACS Observations of RR Lyrae Stars in Six Ultra-deep Fields of M31\altaffilmark{1}}


\author{
Elizabeth J.Jeffery\altaffilmark{2},
Ed Smith\altaffilmark{2},
Thomas M. Brown\altaffilmark{2},
Allen V. Sweigart\altaffilmark{3},
Jason Kalirai\altaffilmark{2},
Henry C. Ferguson\altaffilmark{2},
Puragra Guhathakurta\altaffilmark{4},
Alvio Renzini\altaffilmark{5},
R. Michael Rich\altaffilmark{6}
}

\altaffiltext{1}{Based on observations made with the NASA/ESA {\it Hubble
Space Telescope}, obtained at STScI, which
is operated by AURA, Inc., under NASA contract NAS 5-26555.}

\altaffiltext{2}{Space Telescope Science Institute, 3700 San Martin Drive,
Baltimore, MD 21218;
jeffery@stsci.edu,
edsmith@stsci.edu,
tbrown@stsci.edu,
jkalirai@stsci.edu,
ferguson@stsci.edu,
}

\altaffiltext{3}{Code 667, NASA Goddard Space Flight Center, 
Greenbelt, MD 20771; allen.v.sweigart@nasa.gov}
\altaffiltext{4}{University of California Observatories/Lick Observatory, 
Department of Astronomy \& Astrophysics, 
University of California, Santa Cruz, CA 95064; raja@ucolick.edu}
\altaffiltext{5}{Osservatorio Astronomico, Vicolo Dell'Osservatorio 5,
I-35122 Padova, Italy; alvio.renzini@oapd.inaf.it}
\altaffiltext{6}{Department of Physics and Astronomy, 
University of California, Los Angeles, CA 90095; rmr@astro.ucla.edu}

%
\begin{abstract}

   We present HST/ACS observations of RR Lyrae variable stars in six ultra 
deep fields of the Andromeda galaxy (M31), including parts of the halo, disk, 
and giant stellar stream.  Past work on the RR Lyrae stars in M31 has focused on 
various aspects of the stellar populations that make up the galaxy's halo, 
including their distances and metallicities.  This study builds upon this 
previous work by increasing the spatial coverage (something that has been 
lacking in previous studies) and by searching for these variable stars in 
constituents of the galaxy not yet explored.  Besides the 55 RR Lyrae stars we 
found in our initial field located 11kpc from the galactic nucleus, we find 
additional RR Lyrae stars in four of the remaining five ultra deep fields as 
follows: 21 in the disk, 24 in the giant stellar stream, 3 in the halo field 
21kpc from the galactic nucleus, and 5 in one of the halo fields at 35kpc.  No 
RR Lyrae were found in the second halo field at 35kpc.  The RR Lyrae populations 
of these fields appear to mostly be of Oosterhoff I type, although the 11kpc 
field appears to be intermediate or mixed.  We will discuss the properties of these stars 
including period and reddening distributions.  We calculate metallicities and 
distances for the stars in each of these fields using different methods and 
compare the results, to an extent that has not yet been done.  We compare 
these methods not just on RR Lyrae stars in our M31 fields, but also on a data set 
of Milky Way field RR Lyrae stars.

\end{abstract}


\section{Introduction}
   \label{intro}

   RR Lyrae variable stars have long been an important and useful tool for 
studying stellar populations.  They are relatively easy to identify, given 
their narrow range of $V$ magnitudes, short periods, and distinctive 
light curve shapes (at least for the RR$ab$ type).   Their very presence is 
indicative of an old ($\gtrsim$ 10 Gyr) stellar population and their pulsation 
properties (namely amplitudes and periods) can be used to find the 
metallicities and distances of their parent population.

\begin{table*}
  \begin{center}
  \begin{tabular}{lccccc}

    \hline
                           & Dates  & R.A. & Dec.  & n$_{exp}$ $\times$ total (ks) & n$_{exp}$ $\times$ total (ks)\\
 \multicolumn{1}{c}{Field} & Observed & (J2000) & (J2000)  & (F606W) & (F814W) \\
    \hline


 disk    & 2004 Dec 11\---2005 Jan 18 & 00 49 08.6 & 42 45 02 & 41 $\times$ 52.8 & 62 $\times$ 78.1 \\
 stream  & 2004 Aug 30\---2004 Oct  4 & 00 44 18.2 & 39 47 32 & 41 $\times$ 52.8 & 62 $\times$ 78.1 \\
 halo11  & 2002 Dec  2\---2003 Jan  1 & 00 46 07.1 & 40 42 39 & 108 $\times$ 138.6 & 126 $\times$ 161.3 \\
 halo21  & 2006 Aug  9\---2006 Aug 28 & 00 49 05.1 & 40 17 32 & 24 $\times$ 28.7 & 40 $\times$ 47.8 \\
 halo35b & 2006 Oct 18\---2007 Jan  6 & 00 54 08.5 & 39 47 26 & 24 $\times$ 28.1 & 44 $\times$ 51.6 \\


   \hline
   \end{tabular}
 \end{center}
   \caption{Coordinates of the center of each observed field in M31, along
        with total number of exposures and combined exposure times in each
        filter.  Units for right ascension are hours, minutes, seconds; units
        for declination are degrees, arcminutes, and arcseconds.}
      \label{obslog}
\end{table*}

   Work on the RR Lyraes in the M31 system started in 1987 when Pritchet \& 
van den Bergh used ground-based observations from the Canada-France-Hawaii 
Telescope to observe a field 9kpc from the galactic center.  They identified 
30 such stars with an estimated completeness of 25\%.  This same field was 
later observed by Dolphin et al. (2004) using the WIYN 3.5m at Kitt Peak. They  
found 24 RR Lyrae stars with an estimated completeness of $\sim$24\%.  Additional work 
has been done on RR Lyraes in the M31 spheroid (Sarajedini et al. 2009), 
M31 satellite galaxies M32 (Alonso-Garcia et al. 2004; Fiorentino et 
al. 2010) and M33 (Sarajedini et al. 2006), as well as some of the globular 
cluster systems in M31 (Clementini et al. 2001; Contreras et al. 2008).

   A comprehensive analysis of M31 RR Lyrae stars was enabled by ultra-deep 
imaging of a single field in the M31 halo, 11 kpc (51') from the nucleus 
(Brown et al. 2004, hereafter Paper I).  In this study they utilized the 
excellent time coverage (250 exposures over 41 days, combining to $\sim$84 
hours of imaging time) afforded by a set of observations taken with the Wide 
Field Camera (WFC) channel of the Advanced Camera for Surveys (ACS) onboard 
the Hubble Space Telescope (HST).  These data were taken to study the star 
formation history of this field by observing individual stars below the main 
sequence turn off (Brown et al. 2003).  These observations, taken in both 
F606W (broad $V$) and F814W ($I$), did that, and they were able to construct a 
color magnitude diagram (CMD) of this field to an unprecedented depth ($V \sim$ 
30.7) with 100\% completeness in the magnitude range of the RR Lyrae stars.

   In this study we build upon and expand the work of Paper I. We have 
observed five more ultra deep fields in various regions of Andromeda, making 
six fields in total. These include observations of the disk, giant stellar 
stream, and several halo fields at varying distances from the galactic center 
(namely one at 21 kpc and two at 35 kpc), in addition to the halo field at 11 
kpc. The completeness for star detection is $\sim$100\% at the magnitude level 
of the RR Lyrae stars.  The identification of RR Lyrae stars within the 
stellar sample of each field is also $\sim$100\%, due to the very high 
signal-to-noise ratio and excellent time sampling of the observations.  In 
fact, it is worth noting that M31 RR Lyrae studies that employ HST 
observations with far less temporal coverage do not generally suffer from 
significant completeness issues (e.g., Sarajedini et al. 2009).  As in Paper 
I, the observations we describe here greatly surpass the requirements of a 
study strictly focused on M31 RR Lyrae stars, because the primary purpose of 
the observations was to investigate the star formation history in various 
fields of Andromeda.  To that end, each region was observed with a long series 
of exposures to obtain photometry below the main sequence turn off (see Brown 
et al. 2006; 2007; 2008 for those results). 


   In the remainder of this paper, we will proceed as follows.  We present a 
summary of the observations and data reductions in Section \ref{obs}.  In 
Section \ref{variables} we discuss our techniques for searching for the 
variable stars in the images, including identification of the RR Lyraes, as 
well as period determination and light curve fitting.  We will discuss the 
properties of the samples of RR Lyrae stars in Section \ref{prop}, including 
period distributions and Oosterhoff types.  We discuss and compare various 
techniques to determine metallicities and distances in Section 
\ref{metallicity}.  We summarize findings and final conclusions in Section 
\ref{conclusion}.

\section{Observations and Data Reduction}
  \label{obs}

\begin{figure*}
   \epsscale{0.75}
        \plotone{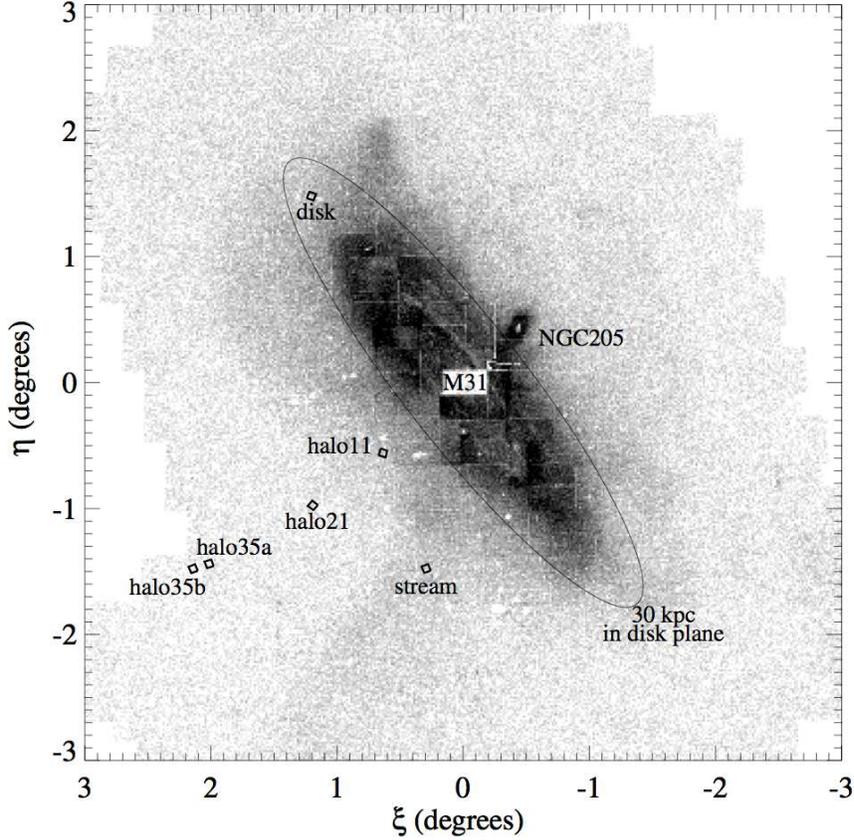}
        \caption{A map of stellar density (from Ferguson et al. 2002) in the
        vicinity of M31 with our six fields indicated.  The ellipse marks the
        area within 30 kpc of the galactic center in the inclined plane (as
        labeled).  Figure taken from Brown et al. (2009). }
   \label{m31fields}
\end{figure*}

   As mentioned in Section \ref{intro}, we obtained the data with HST/ACS with 
the primary goal of exploring the star formation histories of six fields in 
M31 by observing to an unprecedented depth, resolving stars below the main 
sequence turn off (see Brown et al. 2003; 2006; 2007; 2008).  In Figure 
\ref{m31fields} we present the location of these fields on a map of the 
stellar densities in the vicinity of M31 (Ferguson et al. 2002).  (We note 
that although the RR Lyraes in the 11kpc field have been discussed in Paper I, 
we include them in our analysis for completeness.)  These observations 
represent over 300 HST orbits and 220.25 hours of imaging.  The observations of 
each field (including central coordinate information and the number of 
individual exposures for each field) are summarized in Table \ref{obslog}.  A 
more extensive description of these observations is given in Brown et al. 
(2009) and references therein.


   A complete description of our data reduction process to search for variable 
stars is given in Paper I.  In addition to the reductions and corrections 
described there, we also included corrections for charge transfer 
(in)efficiency (CTI, Chiaberge et al. 2009), as well as updated ``breathing'' 
corrections. 

   The halo11, stream and disk observations were performed early in the 
lifetime of the ACS/WFC CCD detector and those data were shown to be unaffected 
by CTI.  The deep field photometry of the halo21 and both halo35 fields (Brown 
et al. 2008) included the best CTI corrections available at that time (Reiss 
\& Mack 2004).  Our photometry of the variables in those fields now employs 
the 2009 update to the CTI corrections (Chiaberge et al. 2009). For stars as 
bright as horizontal branch (HB) RR Lyrae in our fields, the CTI correction 
ranges from near 0 to $\sim$ 0.02 magnitudes.

   Furthermore, the reductions now include a breathing correction $\--$  
that is, an exposure by exposure correction for variation in focus of HST 
caused by orbital temperature variations (Makidon et al. 2006).  We remove 
that variation's effect on our aperture photometry, with an exposure by 
exposure normalization factor derived from the breathing induced variation in 
the mean brightness of a large number of red giant branch stars at $\sim$ HB 
brightness.

\section{Characterization of the Variable Stars}
   \label{variables}

\subsection{Finding Variables}
   \label{find}

   Our methods for searching for the variable stars in the fields is similar
to that laid out in Paper I (see Section 3), with slight modification.  To 
summarize, we first restricted our analysis to those stars brighter than
$m_{F606W}$ = 28.28, to include only stars with a signal-to-noise ratio 
$\textgreater$ 5.  Only stars with at least 20 observations in F814W or 12 in
F606W were included.   We then computed the Lomb-Scargle periodogram 
(Lomb 1976; Scargle 1982) for each star in the catalog that met these criteria 
(regardless of position on the CMD).  This algorithm searches for periodic 
signals in irregularly sampled data and quantifies its statistical
significance.  We accepted only stars with a score of less than 0.01, meaning 
that the chance that this signal arose from random noise is less than 1\%.  We 
relaxed this threshold to 2.5\% for the two halo35 fields.  We did not attempt 
to find stars again in the halo11 field.

   As a second pass, we normalize the photometry of both bands by subtracting 
their median magnitude and then scale the F606W values by 0.7, the approximate 
I/V ratio of RR Lyrae amplitudes.  We then combine the scaled F606W values 
with F814W.  This improves the time sampling in order to find additional RR 
Lyrae stars that may have been missed during the first pass.  This typically 
yielded no more than one or two more RR Lyrae stars in a given field.


   We found RR Lyraes in five of the six ultra deep fields examined.  As noted 
in Paper I, there are 55 in the halo11 field.  In addition to these, we 
found RR Lyrae stars in the other fields as follows: 21 in the disk, 24 in the 
stream, 3 in halo22, none in halo35a, and five in halo35b.  We note that we 
found no RR Lyrae stars in the halo35a field.  This was not a surprise, given 
the low stellar density (roughly 1:3) compared to the halo21 field where we 
found three RR Lyrae stars.  We may have expected to find a similar number of 
RR Lyrae stars in the halo35a field as compared to the halo35b field (where we 
found five RR Lyraes), as the two fields have nearly equal stellar density.  
Thus we thoroughly re-examined our processing and variable finding analysis of 
the halo35a data.  Still finding no RR Lyrae, we added an additional search 
method.  We processed the photometry of all, $\sim$60, stars in or near the RR 
Lyrae region of the CMD with the template fitting program (see Section 
\ref{perLC}).  This till yielded null results.  We are therefore confident in 
the lack of RR Lyraes in this field.


   We have included a complete list of the number of RR Lyraes, the ratio of 
c-type to the total number of RR Lyraes in each field, and average periods of 
each type for each field in Table \ref{char_table}.  For comparison, we have 
also listed the corresponding values for the Milky Way (MW) globular clusters 
(GCs) of Oosterhoff (1939) types I and II (from Clement et al. 2001).


\begin{table}[b]
  \begin{center}
  \begin{tabular}{lcccc}

    \hline
   & Total    &  &   &   \\
 \multicolumn{1}{c}{Field} & RR Lyrae & N$_{RR_{c}}$/N$_{RR_{tot}}$ & $<P_{ab}>$  & $<P_{c}>$  \\
    \hline


 disk      & 21  & 0.43  & 0.583  & 0.341 \\
 stream    & 24  & 0.33  & 0.560  & 0.336 \\
 halo11    & 55  & 0.47  & 0.594  & 0.318 \\
 halo21    &  3  & 0.00  & 0.599  & \---  \\
 halo35a   &  0  & \---  & \---   & \---  \\
 halo35b   &  5  & 0.40  & 0.495  & 0.359 \\
 MW GC OoI &\--- & 0.22  & 0.559  & 0.326 \\
 MW GC OoII&\--- & 0.48  & 0.659  & 0.368 \\


   \hline
   \end{tabular}
 \end{center}
   \caption{Summary of the properties of the RR Lyraes in our M31 fields.
                Properties of the Milky Way globular clusters (MW GC) of
                Oosterhoff type I and II (Clement et al. 2001) are included
                for comparison.  The last three columns indicate the ratio of
                RR\textit{c} to total RR Lyrae stars, the average period for
                the ab-type stars, and the average period for the c-type stars.}
      \label{char_table}
\end{table}

\begin{figure*}
   \epsscale{0.70}
        \plotone{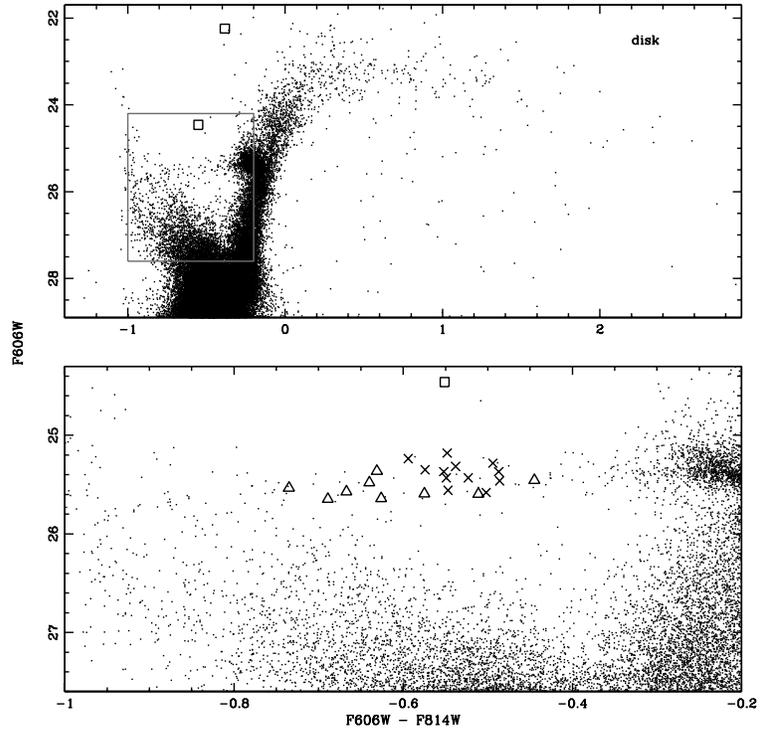}
        \caption{Top: CMD of the M31 disk field from Brown et al. (2009).  The
        Cepheids found in this field are marked as open squares.  The gray box
        indicates the subsection of the CMD displayed in the bottom panel.
        Bottom: Expanded view of the CMD and positions of the
        RR Lyraes on the horizontal branch.  RR\textit{ab} stars
        are indicated with crosses and RR\textit{c} stars with the open
        triangles.}
   \label{cmd_disk}
\end{figure*}

\begin{figure*}
   \epsscale{0.70}
        \plotone{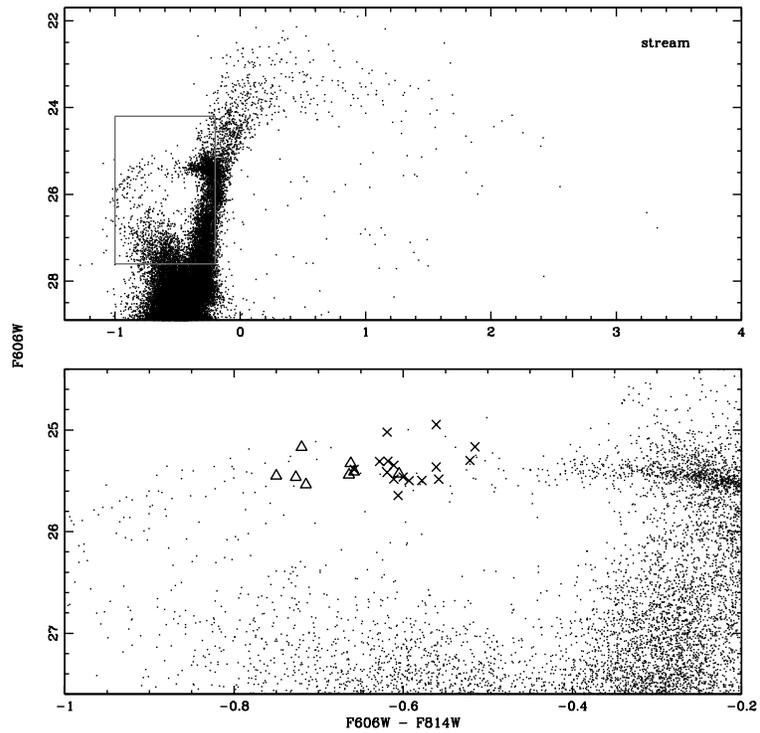}
        \caption{Same as Figure \ref{cmd_disk}, but for the stream field.}
   \label{cmd_strm}
\end{figure*}

\begin{figure*}
   \epsscale{0.70}
        \plotone{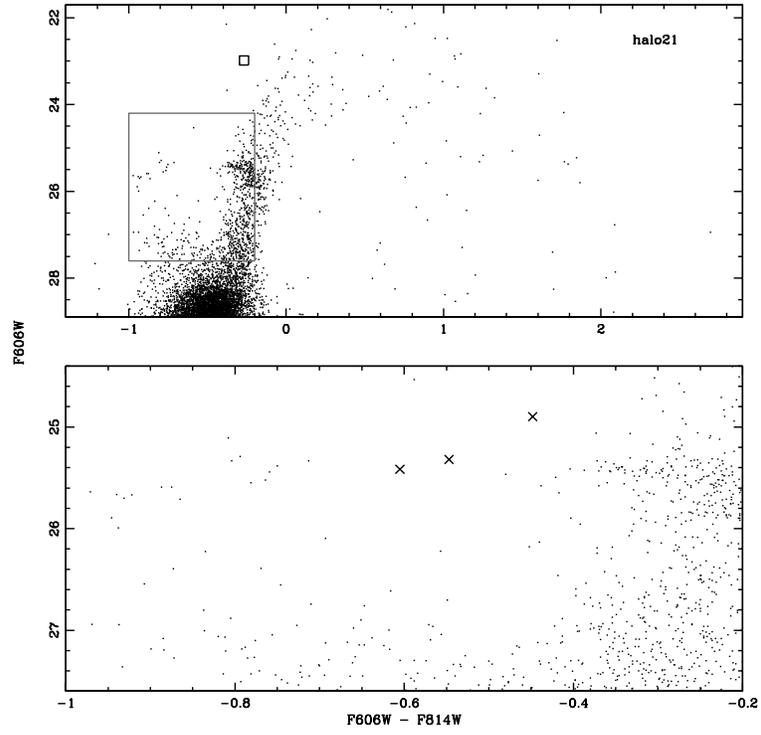}
        \caption{Same as Figure \ref{cmd_disk}, but for the halo21
        field.}
   \label{cmd_halo22}
\end{figure*}

\begin{figure*}
   \epsscale{0.70}
        \plotone{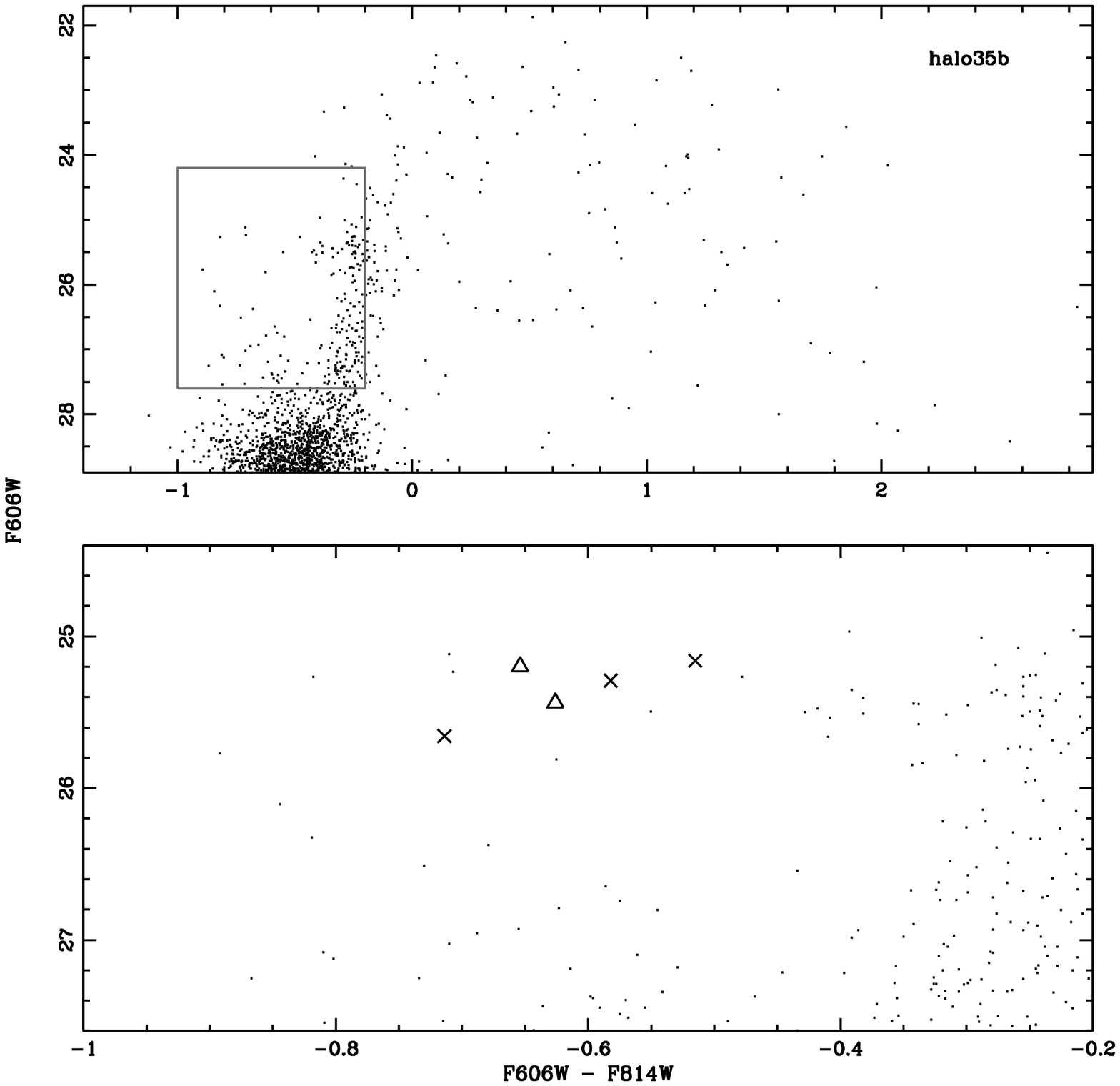}
        \caption{Same as Figure \ref{cmd_disk}, but for the halo35b field.}
   \label{cmd_halo35b}
\end{figure*}

   We indicate the positions of the RR Lyrae stars (and Cepheids, when found, 
see Section \ref{cepheids}) on the CMD of each field in Figures \ref{cmd_disk} 
through \ref{cmd_halo35b}.  We refer the reader to Paper I for the CMD of the 
halo11 field.  The photometry of the CMDs is from Brown et al. (2009).  The 
photometry for the RR Lyraes is based on the average magnitude and color found 
by light curve template fitting (as explained below).  In these figures, the 
top panel represents the CMD of the entire field, while the lower panel is a 
region zoomed to include just the RR Lyrae region on the horizontal branch (as 
indicated by the gray box in the top panel).  In these figures, we represent 
the ab-type RR Lyraes with crosses, while the c-type variables are open 
triangles.


   To help us understand the reliability and completeness of our finding 
routine, we created artificial light curves to process and determine the 
fraction that were successfully recovered.  These synthetic light curves had 
average F606W and F814W magnitudes comparable to our real data and were 
created with a combination of all periods between 0.2 and 1.0 days (in 
increments of 0.1 days), and all F814W amplitudes between 0.1 and 1.5 
magnitudes (in increments of 0.1, and scaling the corresponding synthetic 
F606W amplitudes accordingly).  We then sampled these synthetic light curves 
according to the time sampling of our fields.  We used the modified Julian 
date (MJD) values for the stream and the halo21 fields, and expect similar 
results for the disk and both halo35 fields, respectively, as the time 
sampling is similar.  We then scattered individual synthetic photometry values 
with random Gaussian errors, according to the errors of the photometry in our 
real data.

   We input these simulated light curves into our finding algorithm.  For time 
sampling similar to the stream and disk fields, 100\% of the synthetic stars 
were recovered.  For the less frequently sampled halo21 and halo35 fields,
97\% were still recovered.  We note that of the 3\% of the stars that were not
recovered in this instance, most had very small amplitudes (0.1 magnitudes).  
Because of this, we are confident that our finding routine is indeed able to 
find all the RR Lyraes that are present in these fields.  

   Once we have a list of candidate RR Lyraes in our fields from our finding 
routine, we are ready to determine exact periods and amplitudes.

\subsection{Periods and Light Curve Fitting}
   \label{perLC}

   Once the RR Lyrae candidates in our M31 fields were identified as described 
in the previous 
section, we determined their periods and other pulsation properties.  Because 
the RR Lyraes in the halo11 field were so well sampled in time, in Paper I we 
used high order polynomials to fit the light curves.  However, as 
can be seen from Table \ref{obslog}, subsequent observations of the other ultra 
deep fields consisted of increasingly sparser observations.  While this did 
not appreciably affect the completeness level of the observations of RR Lyrae stars, it 
did affect the reliability of the polynomial fits.  To mitigate this, we 
employed a light curve template fitting program, specifically designed for RR 
Lyrae stars.

   We used the program FITLC\footnote{\url{http://www.mancone.net/fitlc/index.php}} 
(Sarajedini et al. 2009) to determine periods and fit template light curves to 
the data.  This template-fitting, period-finding algorithm is a reincarnation 
of the FORTRAN code written for the Layden \& Sarajedini (2000) study, 
rewritten using the Interactive Data Language (IDL), incorporating a graphical 
user interface (GUI).  The algorithm uses 10 template light curves: six 
ab-type RR Lyraes, two c-type RR Lyraes, one eclipsing binary, and one contact 
binary.  It searches over a user-specified period range in user-specified 
period increments, looking for the fit that minimizes the $\chi^{2}$ value, 
simultaneously fitting multiple filters, as available. In addition to the 
period, FITLC also returns the amplitude and mean magnitude of the star in 
each filter.

   When running FITLC on an individual RR Lyrae star, we first set the period 
to search over the default period range and increment, namely 0.2 to 1.0 day at 
0.01 day increments.  This returned a good estimate of the period.  From 
there, we refined the search to find a more precise fit by decreasing the 
period range and increment, until a period was found to five decimal places 
(i.e., the period increment was equal to 10$^{-5}$ day).  We note that for 
consistency in our analysis we also fit stars in the halo11 field using FITLC.  
Both the polynomial fits in Paper I and the present template fit gave 
comparable results.  The consistency of the two methods for period fitting is 
indicative of the reliability of FITLC to providing high quality fits for RR 
Lyrae light curves.

\begin{figure*}
   \epsscale{1.00}
        \plotone{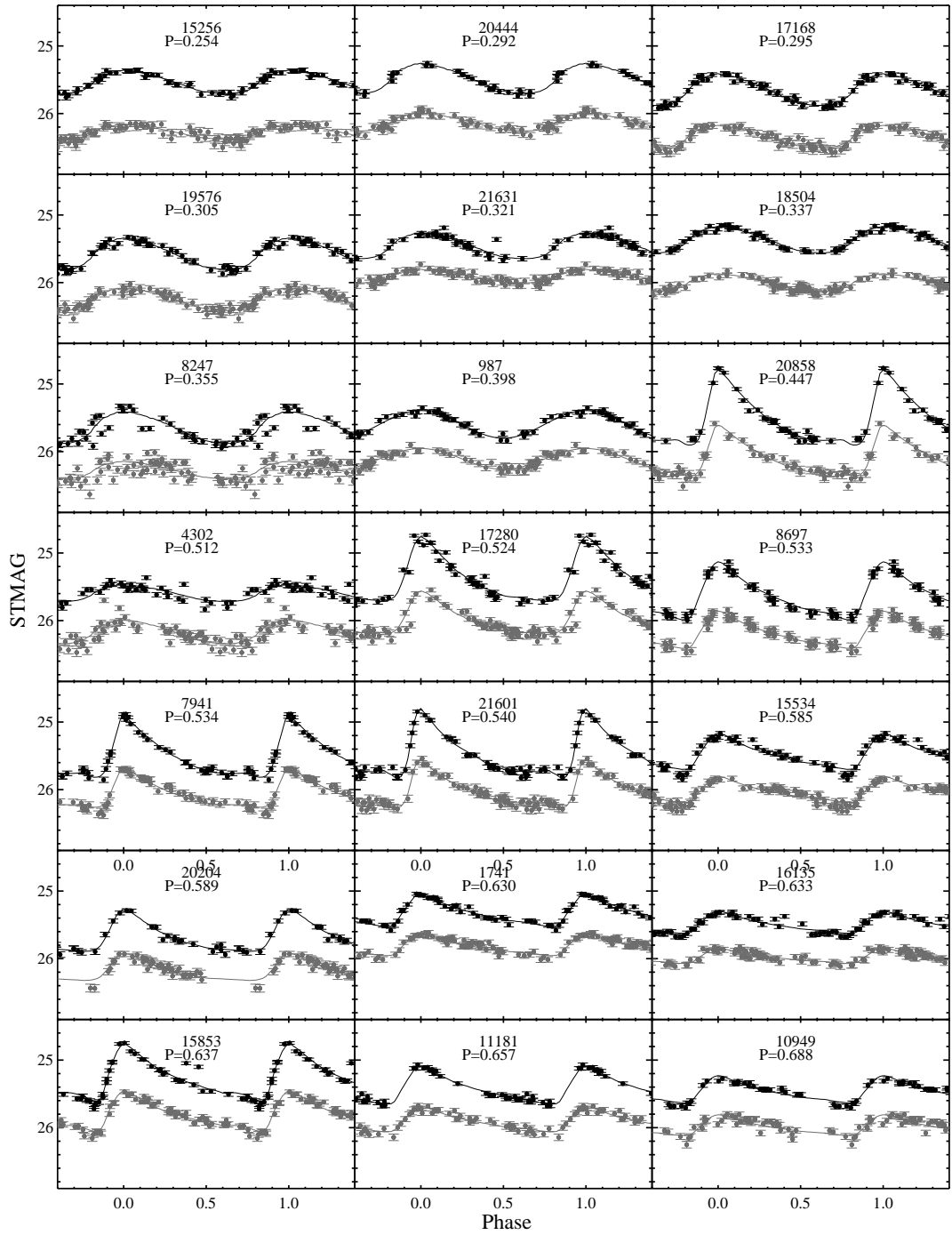}
        \caption{Phased light curves of RR Lyrae stars in the disk, arranged in
                  order of increasing period.  Error bars are included, and
                  the best fit RR Lyrae template light curve is also plotted.
                  The dark points are observations in F606W, while the light
                  gray points are in F814W.}
         \label{diskLC}
\end{figure*}

\begin{figure*}
   \epsscale{1.00}
        \plotone{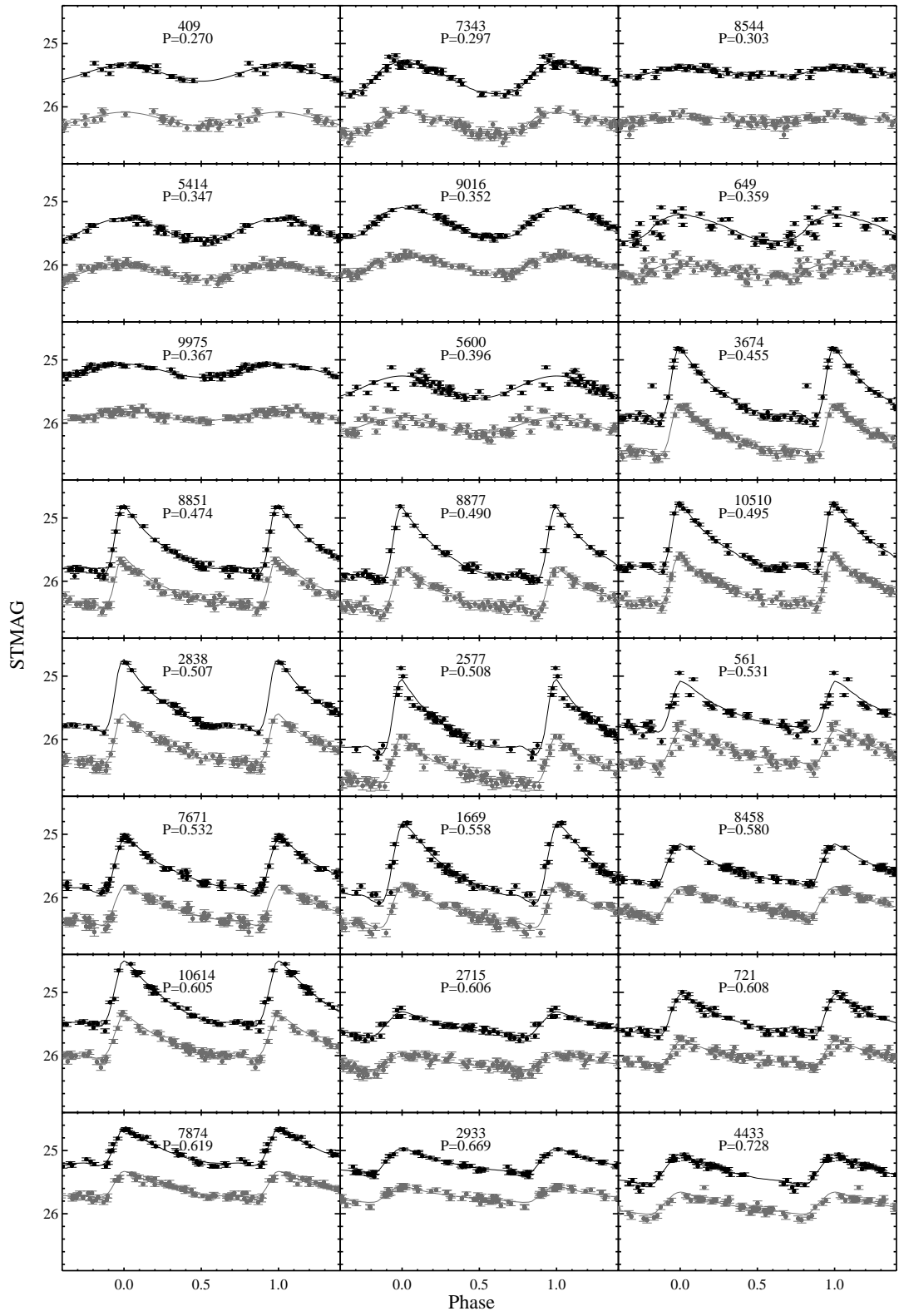}
        \caption{Same as Fig. \ref{diskLC}, but for the stream field RR Lyraes.}
   \label{strmLC}
\end{figure*}

\begin{figure*}
   \epsscale{1.00}
        \plotone{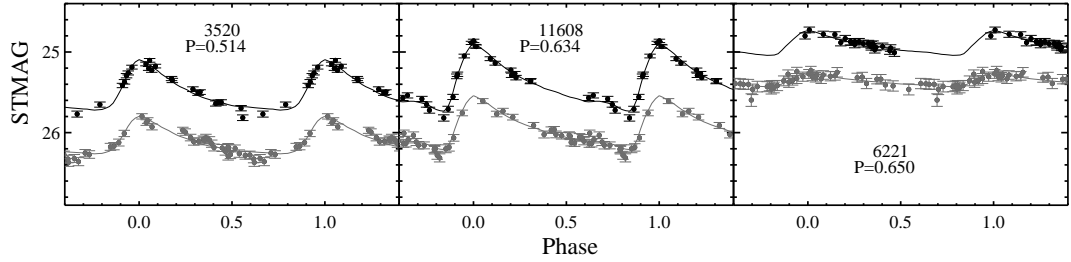}
        \caption{Same as Fig. \ref{diskLC}, but for the halo21 field RR Lyraes.}
   \label{halo21LC}
\end{figure*}

\begin{figure*}
   \epsscale{1.00}
        \plotone{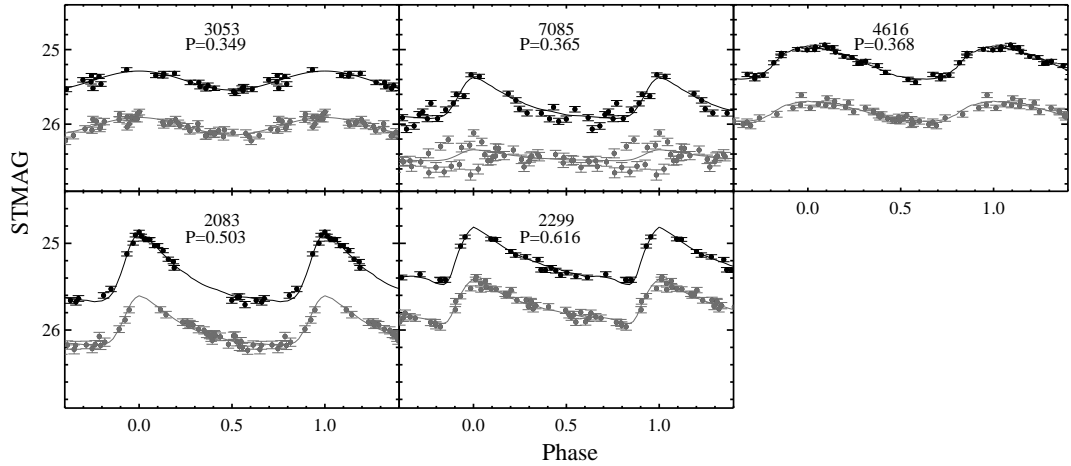}
        \caption{Same as Fig. \ref{diskLC}, but for the halo35b field RR
                Lyraes.}
   \label{halo35bLC}
\end{figure*}

   In Tables \ref{diskOBS} to \ref{35bOBS} we list the individual photometry 
values for the RR Lyraes in each field, excluding the halo11 field.  We refer 
the reader to Paper I for these data.  Tables \ref{RRdisk} to \ref{RRhalo35b} 
list the properties of each RR Lyrae star in each field, including period, 
amplitude, and intensity-weighted mean magnitude for each filter (as 
calculated by FITLC).  Light curves for the individual RR Lyrae stars in each 
field (except for the halo11 field, see Paper I) are displayed in Figures 
\ref{diskLC} through \ref{halo35bLC}, in order of increasing period with the 
best fit template light curve overplotted.  Error bars are also plotted.  The 
dark gray points are observations in F606W while the light gray points are 
F814W.


\begin{table}
  \begin{center}
  \begin{tabular}{ccccccc}

    \hline
 \multicolumn{1}{c}{MJD} & Filter & 987 & 987 (err) & 1741 & 1741 (err) & $\ldots$ \\
    \hline


  53350.176 & F814W & 26.213 & 0.045 & 25.708 & 0.031 & $\ldots$ \\
  53350.192 & F814W & 26.227 & 0.047 & 25.678 & 0.032 & $\ldots$ \\
  53350.236 & F814W & 26.163 & 0.044 & 25.822 & 0.034 & $\ldots$ \\
  $\vdots$ & $\vdots$ & $\vdots$ & $\vdots$ & $\vdots$ & $\vdots$ & $\vdots$ \\


   \hline
   \end{tabular}
 \end{center}
   \caption{Photometry of RR Lyrae stars in the disk.  The modified Julian day
        (MJD) is for the middle of the observation. This table is available
        in its entirety in machine-readable form in the online journal.  A
        portion is shown here for guidance regarding its form and content.}
      \label{diskOBS}
\end{table}


\begin{table}
  \begin{center}
  \begin{tabular}{ccccccc}

    \hline
 \multicolumn{1}{c}{MJD} & Filter & 409 & 409 (err) & 561 & 561 (err) & $\ldots$ \\
    \hline


  53247.604 & F606W & 99.000 & 9.000 & 25.777 & 0.035 & $\ldots$ \\
  53247.620 & F606W & 99.000 & 9.000 & 25.792 & 0.038 & $\ldots$ \\
  53247.668 & F606W & 99.000 & 9.000 & 25.767 & 0.033 & $\ldots$ \\
  $\vdots$ & $\vdots$ & $\vdots$ & $\vdots$ & $\vdots$ & $\vdots$ & $\vdots$ \\


   \hline
   \end{tabular}
 \end{center}
   \caption{Same as Table \ref{diskOBS}, except for the stream.}
      \label{strmOBS}
\end{table}


\begin{table}
  \begin{center}
  \begin{tabular}{ccccccc}

    \hline
 \multicolumn{1}{c}{MJD} & Filter & 3520 & 3520 (err) & 6221 & 6221 (err) & $\ldots$ \\
    \hline


  53956.860 & F606W & 25.508 & 0.030 & 24.883 & 0.041 & $\ldots$ \\
  53956.909 & F606W & 25.639 & 0.032 & 24.873 & 0.039 & $\ldots$ \\
  53956.924 & F606W & 25.625 & 0.033 & 24.871 & 0.039 & $\ldots$ \\
  $\vdots$ & $\vdots$ & $\vdots$ & $\vdots$ & $\vdots$ & $\vdots$ & $\vdots$ \\


   \hline
   \end{tabular}
 \end{center}
   \caption{Same as Table \ref{diskOBS}, except for the halo21 field.}
      \label{21OBS}
\end{table}


\begin{table}
  \begin{center}
  \begin{tabular}{ccccccc}

    \hline
 \multicolumn{1}{c}{MJD} & Filter & 2083 & 2083 (err) & 2299 & 2299 (err) & $\ldots$ \\
    \hline


  54026.517 & F814W & 99.999 & 9.999 & 25.502 & 0.030 & $\ldots$ \\
  54026.533 & F814W & 99.999 & 9.999 & 25.522 & 0.031 & $\ldots$ \\
  54026.584 & F814W & 25.928 & 0.039 & 25.603 & 0.031 & $\ldots$ \\
  $\vdots$ & $\vdots$ & $\vdots$ & $\vdots$ & $\vdots$ & $\vdots$ & $\vdots$ \\


   \hline
   \end{tabular}
 \end{center}
   \caption{Same as Table \ref{diskOBS}, except for the halo35b field.}
      \label{35bOBS}
\end{table}


\begin{table*}
  \begin{center}
  \begin{tabular}{rcccccccc}

    \hline
      & R.A. & Dec. & Period & $m_{F606W}$ & $m_{F814W}$ & $A_{F606W}$ & $A_{F814W}$ & \\
 \multicolumn{1}{c}{Name} & (J2000) & (J2000) & (days) &   (mag)       &   (mag)       &   (mag)     &  (mag)      & \multicolumn{1}{c}{Type} \\
    \hline


  15256 & 00 49 14.96 & 42 44 09.4 & 0.2536 & 25.535 & 26.274 & 0.349 & 0.219 & c  \\
  20444 & 00 49 15.28 & 42 46 25.6 & 0.2920 & 25.479 & 26.127 & 0.447 & 0.266 & c  \\
  17168 & 00 49 13.20 & 42 45 48.4 & 0.2950 & 25.646 & 26.341 & 0.473 & 0.330 & c  \\
  19576 & 00 49 13.13 & 42 46 57.4 & 0.3052 & 25.573 & 26.246 & 0.466 & 0.336 & c  \\
  21631 & 00 49 13.97 & 42 45 10.2 & 0.3212 & 25.457 & 25.908 & 0.383 & 0.200 & c  \\
  18504 & 00 49 13.78 & 42 46 10.1 & 0.3371 & 25.366 & 26.001 & 0.390 & 0.262 & c  \\
   8247 & 00 49 06.63 & 42 44 41.0 & 0.3554 & 25.641 & 26.275 & 0.473 & 0.257 & c  \\
    987 & 00 48 58.78 & 42 45 04.5 & 0.3979 & 25.588 & 26.102 & 0.408 & 0.307 & c  \\
  20858 & 00 49 01.69 & 42 43 44.4 & 0.4471 & 25.476 & 26.082 & 1.109 & 0.747 & ab \\
   4302 & 00 49 03.49 & 42 44 23.3 & 0.5121 & 25.595 & 26.167 & 0.250 & 0.352 & c  \\
  17280 & 00 49 13.00 & 42 45 56.8 & 0.5243 & 25.347 & 25.966 & 0.960 & 0.643 & ab \\
   8697 & 00 49 06.89 & 42 44 46.5 & 0.5330 & 25.607 & 26.173 & 0.857 & 0.550 & ab \\
   7941 & 00 49 07.09 & 42 44 21.5 & 0.5339 & 25.480 & 26.054 & 0.896 & 0.572 & ab \\
  21601 & 00 49 15.38 & 42 44 12.4 & 0.5402 & 25.448 & 26.009 & 1.031 & 0.710 & ab \\
  15534 & 00 49 15.30 & 42 44 08.8 & 0.5846 & 25.470 & 26.026 & 0.535 & 0.386 & ab \\
  20204 & 00 49 17.83 & 42 45 12.9 & 0.5892 & 25.664 & 26.178 & 0.626 & 0.406 & ab \\
   1741 & 00 49 00.12 & 42 44 47.6 & 0.6297 & 25.318 & 25.817 & 0.483 & 0.330 & ab \\
  16135 & 00 49 11.08 & 42 46 15.1 & 0.6334 & 25.489 & 25.977 & 0.330 & 0.226 & ab \\
  15853 & 00 49 09.65 & 42 46 43.9 & 0.6368 & 25.236 & 25.799 & 0.858 & 0.605 & ab \\
  11181 & 00 49 12.19 & 42 43 32.7 & 0.6571 & 25.409 & 25.904 & 0.567 & 0.375 & ab \\
  10949 & 00 49 07.54 & 42 45 26.8 & 0.6880 & 25.462 & 25.988 & 0.403 & 0.309 & ab \\


   \hline
   \end{tabular}
 \end{center}
   \caption{Properties of the RR Lyrae stars found in the disk.}
      \label{RRdisk}
\end{table*}


\begin{table*}
  \begin{center}
  \begin{tabular}{rcccccccc}

    \hline
   & R.A. & Dec. & Period & $m_{F606W}$ & $m_{F814W}$ & $A_{F606W}$ & $A_{F814W}$ & \\
 \multicolumn{1}{c}{Name} & (J2000) & (J2000) & (days) &   (mag)       &   (mag)       &   (mag)     & (mag) & Type \\
    \hline


    409 & 00 44 26.91 & 39 48 40.1 & 0.2702 & 25.462 & 26.190 & 0.257 & 0.216 & c  \\
  \\
   7343 & 00 44 20.90 & 39 47 06.0 & 0.2967 & 25.535 & 26.259 & 0.516 & 0.355 & c  \\
   8544 & 00 44 09.84 & 39 47 35.8 & 0.3030 & 25.453 & 26.204 & 0.134 & 0.108 & c  \\
   5414 & 00 44 16.66 & 39 48 02.3 & 0.3467 & 25.434 & 26.101 & 0.359 & 0.256 & c  \\
   9016 & 00 44 14.07 & 39 47 06.2 & 0.3517 & 25.327 & 25.996 & 0.478 & 0.300 & c  \\
    649 & 00 44 18.66 & 39 49 14.6 & 0.3584 & 25.411 & 26.077 & 0.425 & 0.197 & c  \\
   9975 & 00 44 23.53 & 39 46 03.9 & 0.3666 & 25.168 & 25.889 & 0.209 & 0.150 & c  \\
   5600 & 00 44 16.43 & 39 47 60.0 & 0.3955 & 25.423 & 26.031 & 0.350 & 0.260 & c  \\
   3674 & 00 44 22.49 & 39 48 06.3 & 0.4550 & 25.531 & 26.219 & 1.175 & 0.830 & ab \\
   8851 & 00 44 12.10 & 39 47 19.6 & 0.4738 & 25.487 & 26.124 & 1.118 & 0.797 & ab \\
   8877 & 00 44 15.83 & 39 47 01.3 & 0.4899 & 25.561 & 26.217 & 1.177 & 0.699 & ab \\
  10510 & 00 44 10.67 & 39 46 52.8 & 0.4946 & 25.446 & 26.092 & 1.122 & 0.777 & ab \\
   2838 & 00 44 19.31 & 39 48 35.3 & 0.5071 & 25.452 & 26.108 & 1.147 & 0.805 & ab \\
   2577 & 00 44 17.16 & 39 48 49.2 & 0.5081 & 25.789 & 26.432 & 1.186 & 0.732 & ab \\
    561 & 00 44 13.40 & 39 49 40.7 & 0.5314 & 25.550 & 26.175 & 0.805 & 0.559 & ab \\
   7671 & 00 44 18.20 & 39 47 12.5 & 0.5315 & 25.570 & 26.179 & 0.886 & 0.636 & ab \\
   1669 & 00 44 25.27 & 39 48 26.0 & 0.5583 & 25.542 & 26.191 & 1.271 & 0.697 & ab \\
   8458 & 00 44 22.24 & 39 46 39.4 & 0.5803 & 25.532 & 26.100 & 0.665 & 0.469 & ab \\
  10614 & 00 44 09.82 & 39 46 54.6 & 0.6050 & 25.143 & 25.791 & 1.037 & 0.682 & ab \\
   2715 & 00 44 20.02 & 39 48 33.8 & 0.6060 & 25.527 & 26.108 & 0.379 & 0.266 & ab \\
    721 & 00 44 14.81 & 39 49 31.3 & 0.6083 & 25.423 & 25.997 & 0.669 & 0.386 & ab \\
   7874 & 00 44 22.06 & 39 46 51.3 & 0.6192 & 25.028 & 25.597 & 0.586 & 0.412 & ab \\
   2933 & 00 44 26.76 & 39 47 59.2 & 0.6694 & 25.196 & 25.714 & 0.399 & 0.253 & ab \\
   4433 & 00 44 23.96 & 39 47 46.7 & 0.7277 & 25.334 & 25.859 & 0.472 & 0.349 & ab \\


   \hline
   \end{tabular}
 \end{center}
   \caption{Properties of the RR Lyrae stars found in the stream.}
      \label{RRstrm}
\end{table*}


\begin{table*}
  \begin{center}
  \begin{tabular}{rcccccccc}

    \hline
     & R.A. & Dec. & Period & $m_{F606W}$ & $m_{F814W}$ & $A_{F606W}$ & $A_{F814W}$ & \\
 \multicolumn{1}{c}{Name} & (J2000) & (J2000) & (days) &   (mag)       &   (mag)       &   (mag)     & (mag) & Type \\
    \hline

  V130 & 00 46 11.91 & 40 42 45.1 & 0.2628 & 25.638 & 26.365 & 0.447 & 0.317 & c  \\
   V89 & 00 46 06.68 & 40 43 21.9 & 0.2670 & 25.523 & 26.258 & 0.412 & 0.291 & c  \\
   V76 & 00 46 05.86 & 40 42 52.9 & 0.2743 & 25.422 & 26.140 & 0.148 & 0.114 & c  \\
  V100 & 00 46 06.94 & 40 43 58.7 & 0.2746 & 25.579 & 26.281 & 0.465 & 0.311 & c  \\
  V137 & 00 46 11.13 & 40 43 48.8 & 0.2804 & 25.595 & 26.280 & 0.337 & 0.218 & c  \\
  V120 & 00 46 12.78 & 40 41 24.6 & 0.2830 & 25.381 & 26.104 & 0.485 & 0.324 & c  \\
   V27 & 00 45 57.64 & 40 43 33.5 & 0.2870 & 25.639 & 26.314 & 0.475 & 0.286 & c  \\
   V40 & 00 46 01.20 & 40 42 19.8 & 0.2886 & 25.396 & 26.139 & 0.162 & 0.100 & c  \\
  V102 & 00 46 08.57 & 40 43 10.5 & 0.3008 & 25.504 & 26.208 & 0.446 & 0.288 & c  \\
   V37 & 00 46 03.26 & 40 40 39.9 & 0.3012 & 25.473 & 26.150 & 0.452 & 0.315 & c  \\
    V8 & 00 46 01.02 & 40 40 44.1 & 0.3055 & 25.603 & 26.264 & 0.362 & 0.285 & c  \\
  V163 & 00 46 15.92 & 40 42 37.9 & 0.3126 & 25.342 & 26.008 & 0.359 & 0.200 & c  \\
   V80 & 00 46 05.78 & 40 43 28.8 & 0.3135 & 25.500 & 26.208 & 0.426 & 0.288 & c  \\
  V131 & 00 46 13.72 & 40 41 30.6 & 0.3266 & 25.498 & 26.143 & 0.373 & 0.219 & c  \\
  V157 & 00 46 16.21 & 40 41 47.6 & 0.3291 & 25.462 & 26.160 & 0.409 & 0.287 & c  \\
  V161 & 00 46 13.32 & 40 44 18.7 & 0.3301 & 25.550 & 26.180 & 0.343 & 0.271 & c  \\
   V11 & 00 45 56.88 & 40 43 44.1 & 0.3307 & 25.619 & 26.215 & 0.418 & 0.322 & c  \\
   V43 & 00 46 01.74 & 40 42 20.6 & 0.3379 & 25.361 & 26.032 & 0.507 & 0.319 & c  \\
    V5 & 00 46 00.61 & 40 40 52.4 & 0.3386 & 25.543 & 26.166 & 0.383 & 0.310 & c  \\
   V59 & 00 46 03.80 & 40 42 36.9 & 0.3388 & 25.450 & 26.111 & 0.425 & 0.286 & c  \\
   V83 & 00 46 08.84 & 40 41 35.5 & 0.3515 & 25.336 & 26.026 & 0.491 & 0.320 & c  \\
   V90 & 00 46 07.71 & 40 42 51.1 & 0.3533 & 25.495 & 26.128 & 0.434 & 0.233 & d  \\
   V95 & 00 46 09.60 & 40 41 39.8 & 0.3616 & 25.409 & 26.046 & 0.363 & 0.242 & c  \\
   V54 & 00 46 04.33 & 40 41 35.2 & 0.3659 & 25.402 & 26.053 & 0.390 & 0.273 & c  \\
   V50 & 00 46 02.68 & 40 42 31.0 & 0.3660 & 25.457 & 26.119 & 0.469 & 0.287 & c  \\
    V1 & 00 45 59.76 & 40 41 18.2 & 0.3816 & 25.364 & 25.989 & 0.377 & 0.276 & c  \\
  V147 & 00 46 11.66 & 40 44 23.3 & 0.4417 & 25.462 & 26.090 & 1.044 & 0.633 & ab \\
   V44 & 00 45 59.93 & 40 43 38.9 & 0.4641 & 25.625 & 26.258 & 1.036 & 0.729 & ab \\
   V47 & 00 46 01.55 & 40 42 44.8 & 0.4958 & 25.519 & 26.120 & 1.083 & 0.757 & ab \\
   V88 & 00 46 09.86 & 40 41 03.5 & 0.5056 & 25.517 & 26.125 & 1.049 & 0.770 & ab \\
   V79 & 00 46 04.60 & 40 44 09.8 & 0.5289 & 25.456 & 26.039 & 1.102 & 0.773 & ab \\
%
%
%
%
%
%
%
   V28 & 00 46 00.08 & 40 42 02.9 & 0.5322 & 25.559 & 26.118 & 0.846 & 0.621 & ab \\
  V162 & 00 46 16.08 & 40 42 25.4 & 0.5332 & 25.535 & 26.124 & 0.919 & 0.668 & ab \\
  V126 & 00 46 10.38 & 40 43 32.5 & 0.5337 & 25.433 & 26.034 & 0.941 & 0.583 & ab \\
   V42 & 00 46 00.77 & 40 43 00.9 & 0.5519 & 25.308 & 25.884 & 1.040 & 0.731 & ab \\
   V57 & 00 46 02.60 & 40 43 10.6 & 0.5544 & 25.379 & 25.978 & 1.021 & 0.687 & ab \\
  V142 & 00 46 11.47 & 40 44 04.1 & 0.5545 & 25.378 & 25.958 & 0.913 & 0.636 & ab \\
  V140 & 00 46 14.51 & 40 41 37.0 & 0.5587 & 25.325 & 25.935 & 1.092 & 0.820 & ab \\
  V133 & 00 46 13.43 & 40 41 54.5 & 0.5717 & 25.339 & 25.927 & 0.859 & 0.646 & ab \\
  V112 & 00 46 11.88 & 40 41 21.5 & 0.5729 & 25.374 & 25.973 & 0.941 & 0.572 & ab \\
  V164 & 00 46 14.21 & 40 43 55.8 & 0.5802 & 25.429 & 25.991 & 0.833 & 0.571 & ab \\
  V166 & 00 46 13.71 & 40 44 24.4 & 0.5825 & 25.443 & 25.978 & 0.665 & 0.550 & ab \\
  V122 & 00 46 13.10 & 40 41 14.2 & 0.5886 & 25.319 & 25.946 & 0.810 & 0.478 & ab \\
  V160 & 00 46 14.30 & 40 43 22.7 & 0.6115 & 25.245 & 25.802 & 0.722 & 0.472 & ab \\
  V167 & 00 46 10.38 & 40 43 44.0 & 0.6214 & 25.402 & 25.941 & 0.603 & 0.413 & ab \\
  V124 & 00 46 09.88 & 40 43 34.7 & 0.6265 & 25.342 & 25.876 & 0.703 & 0.530 & ab \\
   V36 & 00 46 02.04 & 40 41 29.0 & 0.6274 & 25.280 & 25.887 & 1.017 & 0.656 & ab \\
  V123 & 00 46 12.46 & 40 41 42.2 & 0.6314 & 25.254 & 25.813 & 0.659 & 0.478 & ab \\
  V136 & 00 46 10.23 & 40 44 23.2 & 0.6342 & 25.543 & 26.064 & 0.472 & 0.314 & ab \\
   V77 & 00 46 04.22 & 40 44 05.8 & 0.6783 & 25.508 & 26.029 & 0.590 & 0.427 & ab \\
   V10 & 00 46 00.68 & 40 41 02.9 & 0.6872 & 25.138 & 25.717 & 0.931 & 0.662 & ab \\
   V66 & 00 46 04.71 & 40 42 32.6 & 0.7130 & 25.356 & 25.855 & 0.535 & 0.300 & ab \\
  V114 & 00 46 10.11 & 40 42 54.5 & 0.7249 & 25.296 & 25.796 & 0.343 & 0.242 & ab \\
   V82 & 00 46 05.29 & 40 44 02.4 & 0.7350 & 25.457 & 25.991 & 0.527 & 0.334 & ab \\
   V78 & 00 46 08.34 & 40 41 24.6 & 0.7748 & 25.365 & 25.912 & 0.507 & 0.361 & ab \\


   \hline
  \end{tabular}
 \end{center}
   \caption{Properties of RR Lyrae stars found in the halo11 field.}
      \label{RRhalo11}
\end{table*}


\begin{table*}
  \begin{center}
  \begin{tabular}{rcccccccc}

    \hline
     & R.A. & Dec. & Period & $m_{F606W}$ & $m_{F814W}$ & $A_{F606W}$ & $A_{F814W}$ & \\
 \multicolumn{1}{c}{Name} & (J2000) & (J2000) & (days) &   (mag)       &   (mag)       &   (mag)     & (mag) & Type \\
    \hline

   3520 & 00 49 04.97 & 40 19 24.4 & 0.5141 & 25.491 & 26.096 & 0.626 & 0.465 & ab \\
  11608 & 00 49 01.09 & 40 17 47.8 & 0.6341 & 25.367 & 25.924 & 0.843 & 0.668 & ab \\
   6221 & 00 49 11.22 & 40 17 05.4 & 0.6497 & 24.914 & 25.361 & 0.305 & 0.171 & ab \\


   \hline
  \end{tabular}
 \end{center}
   \caption{Properties of RR Lyrae stars found in the halo21 field.}
      \label{RRhalo22}
\end{table*}


\begin{table*}
  \begin{center}
  \begin{tabular}{rcccccccc}

    \hline
     & R.A. & Dec. & Period & $m_{F606W}$ & $m_{F814W}$ & $A_{F606W}$ & $A_{F814W}$ & \\
 \multicolumn{1}{c}{Name} & (J2000) & (J2000) & (days) &   (mag)       &   (mag)       &   (mag)     & (mag) & Type \\
    \hline

   3053 & 00 54 01.43 & 39 46 20.5 & 0.3490 & 25.411 & 26.024 & 0.255 & 0.233 & c  \\
   7085 & 00 54 15.97 & 39 44 18.4 & 0.3650 & 25.724 & 26.439 & 0.551 & 0.148 & ab \\
   4616 & 00 54 07.35 & 39 45 23.3 & 0.3680 & 25.174 & 25.830 & 0.445 & 0.272 & c  \\
   2083 & 00 53 58.01 & 39 46 47.1 & 0.5030 & 25.376 & 25.970 & 0.792 & 0.576 & ab \\
   2299 & 00 53 58.77 & 39 46 41.3 & 0.6164 & 25.193 & 25.711 & 0.663 & 0.512 & ab \\


   \hline
  \end{tabular}
 \end{center}
   \caption{Properties of RR Lyrae stars found in the halo35b field.}
      \label{RRhalo35b}
\end{table*}


\newpage


\begin{figure*}
   \epsscale{1.00}
        \plotone{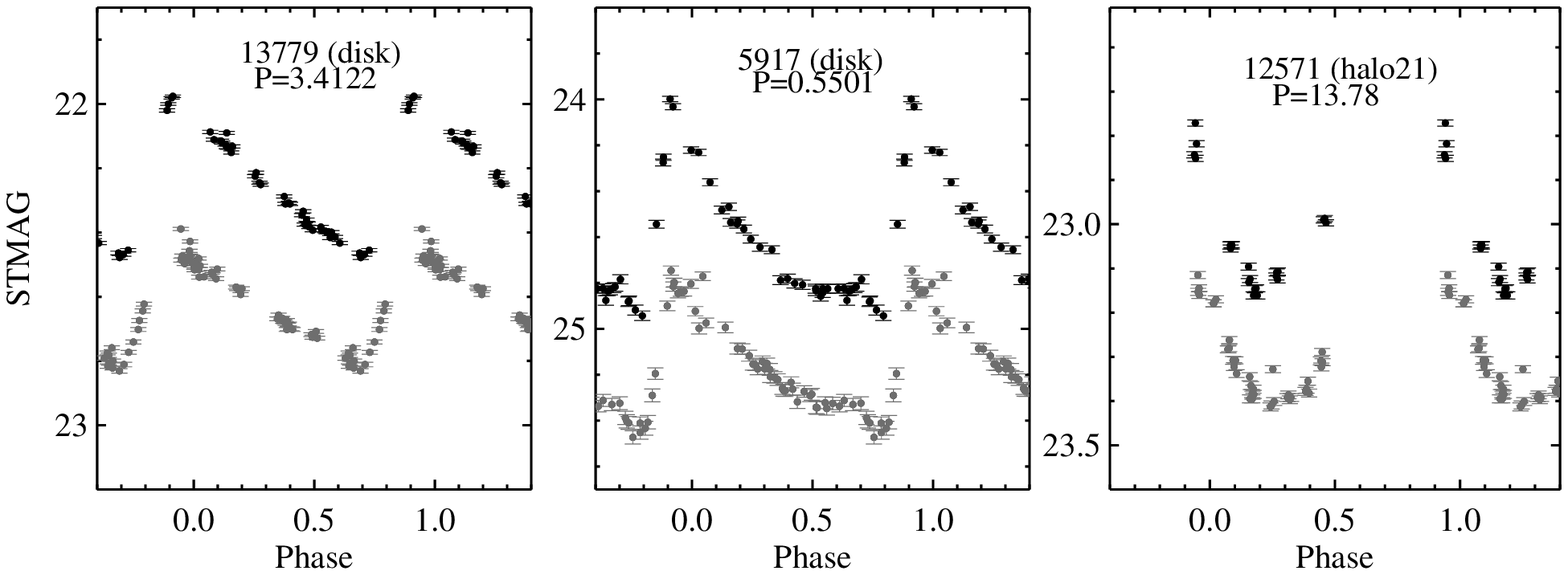}
        \caption{Phased light curves for the three Cepheids found in our
                sample.  Periods (in days) are listed for each star.}
   \label{cepLC}
\end{figure*}

   To test the period and light curve type determination of FITLC, we took the 
same set of synthetic light curves described in Section \ref{find} and 
processed them through FITLC.  The results were excellent; the average period 
difference was less than 0.0001 days.  The light curve type was correctly 
identified for 99.9\% of the stars for ab-type and for 99.3\% of the c-type.  
(We note that we counted an identification as correct as long as any template 
for the correct type provided the lowest $\chi^{2}$ value.)  This reinforces 
our confidence in the results of FITLC.

\subsection{Cepheids}
   \label{cepheids}

   In addition to the RR Lyraes, our search for variable stars has yielded 
three Cepheids: two in the disk (star 13779 and star 5917) and one in the 
halo21 field (star 12571).  We have marked these stars on the appropriate CMDs 
with open squares (see Figures \ref{cmd_disk} and Figure \ref{cmd_halo22}).  
In Tables \ref{other_disk} and \ref{other_strm} we list the individual 
photometry values for these Cepheids.  We plot the phased light curve in both 
F606W and F814W in Figure \ref{cepLC}, 
and note their periods.  The star 5917 in the disk appears to be an 
anomalous Cepheid (Zinn \& Searle 1976), with a (relatively) short period (P 
$\sim$ 0.5501 days) and a magnitude brighter than the RR Lyrae stars.

   We list the properties of these Cepheids, including coordinate information, 
in Table \ref{cep_info}.


\begin{table}[b]
  \begin{center}
  \begin{tabular}{ccccccc}

    \hline
 \multicolumn{1}{c}{MJD} & Filter & 5917 & 5917 (err) & 13779 & 13779 (err) \\
    \hline


  53350.176 & F814W & 25.411 & 0.027 & 22.775 & 0.007 \\
  53350.192 & F814W & 24.799 & 0.019 & 22.782 & 0.007 \\
  53350.236 & F814W & 25.291 & 0.026 & 22.759 & 0.006 \\
  $\vdots$ & $\vdots$ & $\vdots$ & $\vdots$ & $\vdots$ & $\vdots$ \\


   \hline
   \end{tabular}
 \end{center}
   \caption{Photometry of the two Cepheids found in the disk field.  This
        table is available in its entirety in machine-readable form in the
        online journal.  A portion is shown here for guidance regarding its
        form and content.}
      \label{other_disk}
\end{table}


\begin{table}[b]
  \begin{center}
  \begin{tabular}{cccccc}

    \hline
 \multicolumn{1}{c}{MJD} & Filter & 12571 & 12571 (err) \\
    \hline


  53956.860 & F606W & 22.843 & 0.007 \\
  53956.909 & F606W & 22.771 & 0.007 \\
  53956.924 & F606W & 22.851 & 0.007 \\
  $\vdots$ & $\vdots$ & $\vdots$ & $\vdots$  \\


   \hline
   \end{tabular}
 \end{center}
   \caption{Same as Table \ref{other_disk}, but for the Cepheid in the halo21
        field.}
      \label{other_strm}
\end{table}


\begin{table*}
  \begin{center}
  \begin{tabular}{rcccccccc}

    \hline
     & R.A. & Dec. & Period & $m_{F606W}$ & $m_{F814W}$ & $A_{F606W}$ & $A_{F814W}$ & \\
 \multicolumn{1}{c}{Name} & (J2000) & (J2000) & (days) &   (mag)       &   (mag)       &   (mag)     & (mag) & Field \\
    \hline


   5719 & 00 49 04.67 & 42 44 33.4 & 0.5501 & 24.615 & 25.168 & 0.941 & 0.659 & disk  \\
  13779 & 00 49 10.54 & 42 45 25.9 & 3.4122 & 22.245 & 22.628 & 0.502 & 0.383 & disk \\
  12571 & 00 49 06.74 & 40 16 07.0 & 13.78  & 22.985 & 23.252 & 0.359 & 0.323 & halo21  \\


   \hline
   \end{tabular}
 \end{center}
   \caption{Properties of the Cepheids found in our fields.}
      \label{cep_info}
\end{table*}

\section{RR Lyrae Properties}
   \label{prop}

   Here we discuss the various properties of the RR Lyrae population in each 
field.  This includes the period distribution, the Oosterhoff type, and 
reddening.

\subsection{Period Distribution}
   \label{period}

\begin{figure}
   \epsscale{1.30}
        \plotone{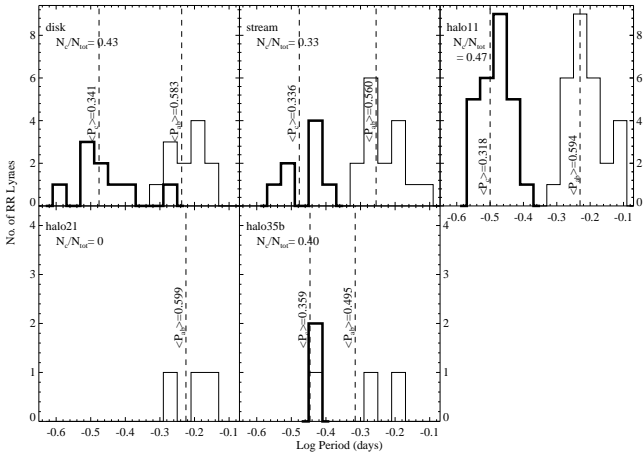}
        \caption{Period distributions of RR Lyrae stars in our five M31 fields
        where these stars were found.  RR\textit{ab} stars are indicated with
        thin lines while the RR\textit{c} stars are the thick lines.  The
        average of the distributions of each type is indicated by the dashed
        vertical lines; the ratio of c-type to total RR Lyraes is also
indicated.}
   \label{period_dist}
\end{figure}

   In Figure \ref{period_dist} we display the period distribution of the stars 
in each of the fields; RR\textit{ab} stars are represented with the thin lines 
while the RR\textit{c} stars are indicated by the thick lines.  As expected, 
the two types of RR Lyrae stars separate from each other.  In this figure we 
have indicated the average periods for both types, and indicated these with the 
vertical dashed lines.

   We note the presence of two seeming outliers in the period distributions 
for their respective type: the star 4302 (a c-type star with an unusually long 
period) in the disk, and the star 7085 (an ab-type star with an unusually 
short period) in the halo35b field.  In the examination of 4302, we noted two 
high points in the F814W band photometry (near phase 0.0 and replotted at 1.0, 
see Figure \ref{diskLC}), leading us to question the c-type fit as correct.  We 
investigated if these two points were indicative of the familiar saw-tooth 
shape of the ab-type.  A check of the star on the individual images did not 
indicate any anomalous cosmic ray hits, hot pixels, or contamination from a 
nearby star.  However, given the higher quality fit to the c-type template 
when compared to the RR\textit{ab} template, and comparing the fit of the 
photometry on the light curve both immediately preceding and following these 
two points, we are convinced that the classification of 4302 as a c-type is 
correct.  This is interesting, given its relatively long period, though we are 
unsure why this particular population would form such a star.  Long 
period c-types are defined as having periods longer than 0.45$^{d}$ (Contreras 
et al. 2005), and only a few of these stars have been found in MW GCs, 
including $\omega$ Cen, NGC 6388, NGC 6441, and M3 (Catelan 2004).  


   It remains unclear as to why some populations can produce long period 
c-types, while others do not.  It is interesting to note that the Galactic 
GCs with long-period RRc stars (listed above), are 
among the most massive GCs in the Milky Way.  Additionally, the CMD morphology 
of these clusters indicates the presence of multiple populations (e.g., Bedin 
et al. 2004; Rich et al. 1997; Piotto et al. 2002).  Analysis of multiple main 
sequences in $\omega$ Cen, for example, show the helium abundance (Y) in the 
cluster may have also changed drastically from one star formation episode to 
the next (Piotto et al. 2005).  Interestingly, in studying different scenarios 
possibly responsible for bimodal horizontal branches in the massive GCs NGC 
6388 and NGC 6441, Yoon et al. (2008) found that the super-helium-rich 
scenario would lead to c-types of longer periods.  (These two particular 
clusters, however, also exhibit other strange properties, including an 
unusually high average period for RR\textit{ab} type stars, and not fitting 
into either Oosterhoff type; see Pritzl et al. 2000 for more details).  
However, tests of the horizontal branch morphology of M3 show no signs of helium 
enrichment (Catelan et al. 2009).  While these explanations are intriguing, it 
remains unclear as to the true explanation for these long period c-types, and 
the significance, if any, of the presence of one in our disk field.

   Second, we note star 7085 (P=0.365$^{d}$) in the halo35b field.  Initially 
we investigated if this star's short period would place it among the c-type 
stars.  However, the ab-type template is a higher quality fit (see especially 
the fit to the F606W data in Figure \ref{halo35bLC}), and we note the existence 
of a handful of ab-type RR Lyraes with similarly short periods in other studies 
(e.g., Sarajedini et al. 2009).  We note the scatter in the F814W band (see 
Figure \ref{halo35bLC}).  Although some data points in the F814W band cover a 
much larger time range than the F606W (i.e., of order 80 days), the exclusion of 
these points does not increase the quality of the fit.  This indicates that the 
scatter is not the result of the Blazhko effect (Blazhko 1907).  This star is 
fainter than the other RR Lyraes in this field, so it could be simply the result 
of increased photometric scatter (although the error bars are small).  Because 
of our limited information we leave the classification as it is, and proceed 
with our analysis.

\subsection{Oosterhoff Type}
   \label{ootype}

   Oosterhoff (1939) noticed similarities among RR Lyraes in different 
globular clusters (GCs) in the Milky Way and subsequently classified these 
variables into two classes, Type I and Type II (often denoted as OoI and OoII, 
respectively).  OoI GCs tend to have a smaller ratio of c-type to total RR 
Lyraes, with shorter average periods, and hence more metal rich compared to 
the OoII GCs.  The two types also occupy different regions of the Bailey 
diagram, a plot of the amplitude vs. the logarithm of the period for 
individual stars.  The so-called Oosterhoff gap is the gap between these two 
types, especially evident in a plot of period shift (shifted relative to M3 RR 
Lyrae stars) versus metallicity (e.g., Suntzeff, Kinman, \& Kraft 1991, see 
especially their Figure 8).  RR Lyraes in the Milky Way field, while more 
difficult to characterize than those in the GCs, seem to show a higher 
tendency toward the OoI type (Cacciari \& Renzini 1976), and still exhibit the 
Oosterhoff gap.

   Recent studies of RR Lyrae stars in dwarf spheroidal (dSph) satellite 
galaxies of the Milky Way reveal that these stars populate an 
intermediate Oosterhoff type, bridging the Oosterhoff gap (Siegel \& 
Majewski 2000).  Similarly, this intermediate Oosterhoff type has been 
observed in the M31 halo (Paper I) and M31 dSph galaxies (e.g., And VI, Pritzl 
et al. 2002)  However, some objects in the M31 system do not show this
intermediate behavior; for example, two spheroid fields of M31 near M32 are of 
OoI type (Sarajedini et al. 2009).

   Initial classification of the stars in these fields as either Oosterhoff I 
or II can be done by examining the average of their period distributions and 
comparing them to those in the MW GCs (see Table \ref{char_table}).  Based on 
this criteria, the RR Lyraes in the stream field most resemble the OoI 
population.  The average periods of the disk stars would lead us to believe 
that the disk may be slightly intermediate with average periods of 
RR\textit{ab} and RR\textit{c} types being slightly higher than OoI, but lower 
than OoII.  (We will see below, however, that this field more closely 
resembles OoI.)  We note that the average period for c-types in the disk is 
higher, due to the presence of star 4302.  Excluding this star, the average 
period for RR\textit{c} stars in the disk is 0.320, in the range of the OoI 
type value.  Likewise, we confirm the results of Paper I of the halo11 field 
being intermediate type.  The numbers of stars in the halo21 and halo35b 
fields are too sparse to meaningfully interpret them, but seem to likely be 
OoI.

\begin{figure}
   \epsscale{1.20}
        \plotone{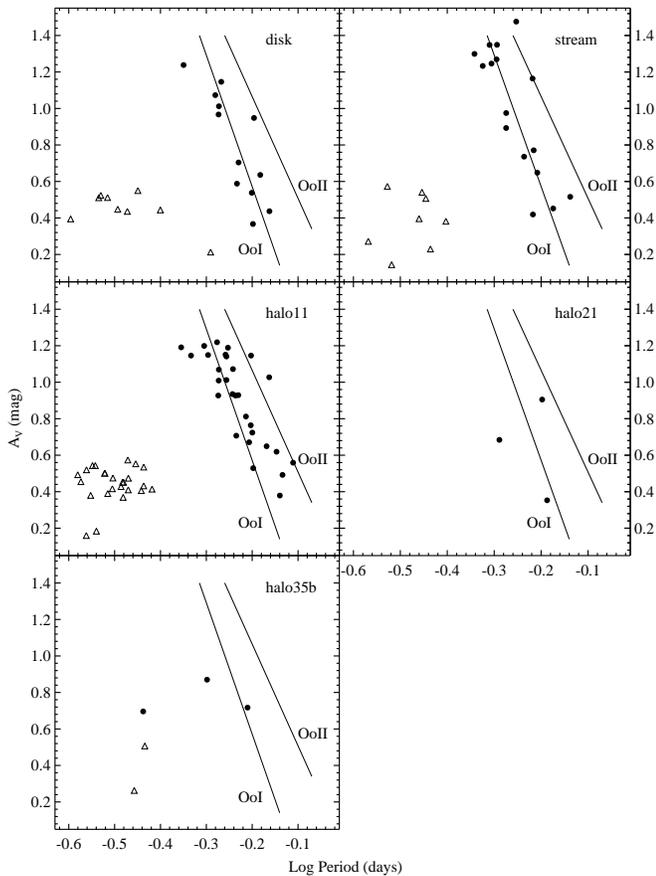}
        \caption{Bailey diagram for the RR Lyrae stars in our fields.  The
                solid lines represent the Oosterhoff classes of the Milky Way
                globular clusters (Clement 2000).  The circles are the
                ab-type stars and the triangles represent the c-type RR
                Lyraes.  The halo11 field appears to be an intermediate case
                between Oosterhoff types. The stream appears to be of OoI type,
                as expected given the ratio of c-type and average period of
                ab-types. The disk also seems to preferentiate to OoI type,
                despite its ratio of c-type RR Lyrae to total RR Lyrae being
                closer to an intermediate type.  Data of the halo21 and
                halo35b fields are too sparse to know for sure which
                Oosterhoff class the populations best resemble, although they
                appear to cluster towards the OoI locus.}
   \label{oo}
\end{figure}

   In Figure \ref{oo} we present Bailey diagrams for the RR Lyrae stars in 
these fields (including the halo11 field, plotting with the updated values as 
listed in Table \ref{RRhalo11}), with the solid lines representing the 
relations for OoI and OoII classes of the Milky Way GCs (Clement 2000).  
For this plot, we converted the observed F606W amplitudes to $V$ band 
amplitudes.  To do this, we took the F606W and F814W photometry from the 
well-sampled, best fit light curve templates for each individual star (see 
Section \ref{perLC}), and applied a color-dependent transformation to convert 
the observed magnitudes to the standard $V$ and $I$ bandpasses.  To perform 
this transformation, we used the synthetic stellar spectra of the MARCS model 
set (Gustafsson et al. 2008), an assumed extinction of $E(B-V)$ = 0.08 
(Schlegel et al. 1998), and the extinction curve of Fitzpatrick (1999).  Once 
this transformation of the light curve in F606W to $V$ was done, we determined 
the amplitude of the pulsations in $V$.


   In this figure, we represent the ab-type stars with circles, and the c-type 
stars with open triangles.  This diagram confirms the intermediate-type 
population of the halo11 field, and that the stream field RR Lyraes are likely 
OoI type.  It also indicates the disk population is most likely OoI type, 
despite its ratio of c-type RR Lyrae to total RR Lyrae stars being 
intermediate of the two types, and its average RR\textit{ab} period slightly 
higher than OoI type.  It is unclear as to why there is this small difference, 
but despite it, we classify the disk as OoI.  While the halo21 and halo35b 
appear to have a tendency toward the OoI type, we emphasize that the statistics 
in these fields are too limited to meaningfully interpret them.

\subsection{RR Lyrae Reddenings}
   \label{red}

   Guldenschuh et al. (2005) showed that the minimum-light color of an 
RR$_{ab}$ star can be used to estimate its line-of-sight reddening.  For 
RR\textit{ab} stars with periods between 0.39 and 0.7 days and metallicities 
ranging from [Fe/H] $\sim \--$3 to 0, they find that intrinsic color depends 
very little on period or metallicity.  They find that the color of these 
variables at minimum-light is $(V-I)_{0,min}$ = 0.58 $\pm$ 0.02.

\begin{figure}
   \epsscale{1.20}
        \plotone{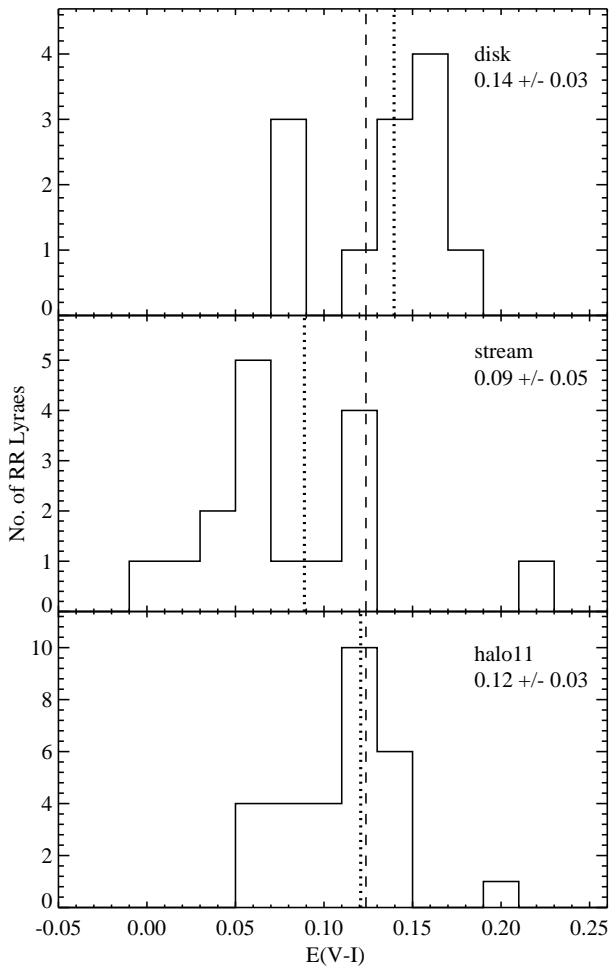}
        \caption{Distributions of $E(V-I)$ for RR$_{ab}$ stars in the three
                fields of our sample with a significant RR Lyrae population.
                The vertical dashed line represents the average line-of-sight
                reddening for M31 (Schlegel et al. 1998), while the vertical
                dotted line is the average of each distribution.  The average
                reddening and standard deviation are indicated on each panel.}
   \label{reddist}
\end{figure}

   Armed with this information, we are able to calculate the reddening of each 
of the RR$_{ab}$ stars in our fields.  To do this, we use the $V$ and $I$ 
magnitudes, as calculated in Section \ref{ootype}.  The observed $(V-I)$ color 
at minimum light was calculated from the best fit light curve template, and 
then used to calculate the reddening, $E(V-I)$, given the intrinsic value.  
We plot the distribution of these calculated reddenings in Figure 
\ref{reddist}.  The vertical dashed line in this figure is the line of sight 
reddening for M31, $E(B-V)$ = 0.08 (Schlegel et al. 1998), converted to 
$E(V-I)$ using the reddening relations of Cardelli, Clayton, \& Mathis (1989). 
In this same figure, the average of each distribution is indicated by the 
vertical dotted line.  This average, along with the standard deviation of the 
distribution is listed on each panel.  This gives us an estimate of the error 
in the reddening of each field. 


   In addition to this error, we explored any error introduced by how well the 
template fitting program (see Section \ref{perLC}) could recover the minimum 
light color of the synthetic light curves described in Section \ref{find}.  The 
mean difference between the synthetic minimum light color and the fit color 
was 0.008 magnitudes, less than the standard deviation of the reddening 
distributions in our fields, and adding little to the previously quoted error.

   We note the peak of the distribution of the halo11 field is at the average 
value for Andromeda, confirming that there is likely little dust adding to the 
reddening in this part of the galaxy.  The stream reddening is less than the 
average value.  It is unclear to us as to why this is, but examination of the 
CMD (see Figure \ref{cmd_strm}) indicates that the RR Lyraes in this field 
are, on average, bluer than those in the other fields.  

   Most of the disk stars fall at somewhat higher reddening than the average 
line-of-sight reddening for M31. This is not surprising given the higher 
amounts of gas and dust in the galactic disk. This higher-than-average 
reddening value may affect the age determined for the disk by the main 
sequence turn off (Brown et al. 2006, see especially their Table 6).  The 
assumption of a larger reddening value would imply the intrinsic population is 
bluer and brighter, which would yield younger ages and lower metallicities 
than those determined by Brown et al. (2006) for the disk field.

\section{Metallicity and Distance}
   \label{metallicity}

   It has been found that the period and other pulsation properties of an RR 
Lyrae star can be used to determine its metallicity (e.g., Alcock et al. 
2000; Sarajedini et al. 2006), and the star's absolute magnitude, and hence 
distance.  Here we discuss various approaches for determining the metallicity 
of the ab-type stars (including the implied distance moduli).  We also review 
a few of the shortcomings of some of the commonly used relations.  After 
discussing and comparing these relations for the ab-type stars, we discuss the 
c-type stars.

\subsection{The Period-Metallicity Relation}
   \label{ata}

   One simple relation for determining RR Lyrae metallicities was derived by 
Sarajedini et al. (2006, hereafter S06) by fitting data from A.C. Layden of 132 
Galactic RR Lyrae stars in the solar neighborhood.  The relation they found is 

\begin{equation}
   [Fe/H] = -3.43 - 7.82 \log{P_{ab}}
\end{equation}
   \label{ataeq}

\noindent
where $P_{ab}$ is the period of the RR$_{ab}$ stars in days.  This relation is 
valid for metallicities between $\--$2.5 and 0.0.  S06 used this relation to 
determine the metallicity distribution function (MDF) of the RR Lyrae stars 
they found in M33.  As they note, this relation is similar in form to that 
determined by Sandage (1993).  Errors for this relation are large (rms $\sim$ 
0.45 dex).  

\begin{figure*}
   \epsscale{1.20}
        \plotone{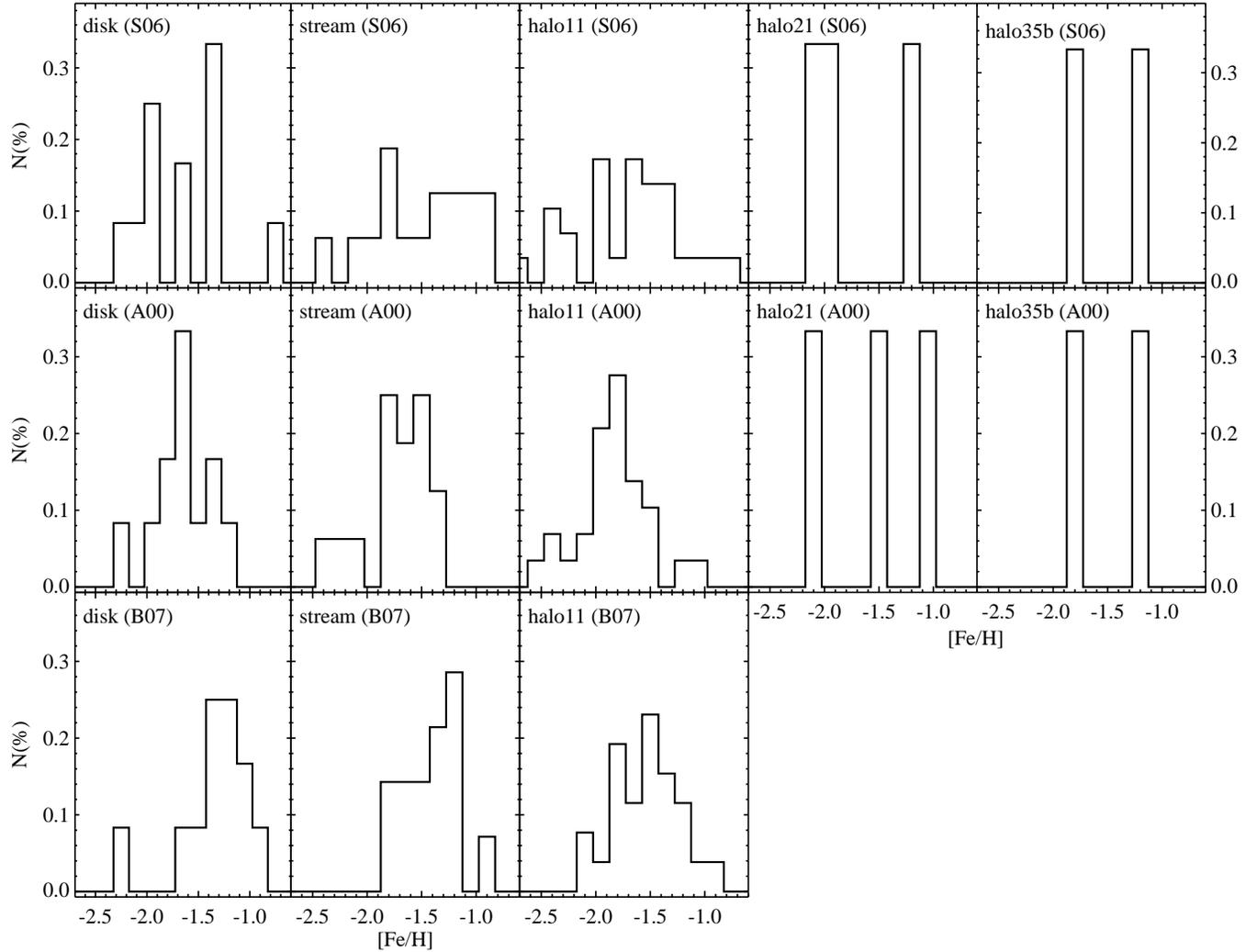}
        \caption{Normalized metallicity distributions for the ab-type RR
                Lyraes stars for each field calculated with the three
                different methods, as labeled.  The panels are tiled in such a
                way for easy comparison of the MDFs produced by each method
                in each field.  We have listed the average, standard
                deviation, and standard error of the mean of each of
                these distributions in Table \ref{calctable}.  We note that
                the MDF of the halo35b field excludes star 7085, although it
                is used when normalizing the MDF.  (See text for discussion).}
   \label{fehdist}
\end{figure*}

   Using this relation, we present the normalized MDF for each of our M31 
fields in the top row of panels of Figure \ref{fehdist}.  We calculate the 
average of each distribution and list these values in Table \ref{calctable}.  
Two errors are quoted for each field: the first is the standard deviation, the 
second (in parenthesis) is the standard error of the mean.  They are 
calculated in the usual way, except for the halo21 and halo35b fields.  In 
these cases we use the formulae appropriate for small statistical samples. 
The standard deviations agree with the expected accuracy of the technique 
($\sim$ 0.45 dex).  


\begin{table*}
  \begin{center}
  \begin{tabular}{cccccc}

    \hline

 [Fe/H]        & disk & stream & halo11 & halo21 & halo35b  \\
    \hline


 S06      & -1.58 $\pm$ 0.42(0.12) & -1.44 $\pm$ 0.45(0.11) & -1.63 $\pm$ 0.46(0.09) & -1.67 $\pm$ 0.47(0.27) & -1.44 $\pm$ 0.61(0.43) \\
 A00      & -1.57 $\pm$ 0.26(0.08) & -1.65 $\pm$ 0.31(0.08) & -1.77 $\pm$ 0.33(0.06) & -1.53 $\pm$ 0.64(0.37) & -1.40 $\pm$ 0.51(0.36) \\
 B07      & -1.29 $\pm$ 0.36(0.10) & -1.32 $\pm$ 0.29(0.07) & -1.47 $\pm$ 0.30(0.06) & \--- & \--- \\
 RR$_{c}$ & -1.70 & -1.65 & -1.43 & \--- &  -1.88 \\


   \hline
   \end{tabular}
 \end{center}
   \caption{Summary of the average calculated metallicities of the RR Lyraes in
                our M31 fields.  The method used is indicated in the first
                column. Errors quoted are the standard deviations of the
                respective distribution, with the standard error of the mean
                in parenthesis.  For the A00 and S06 methods, the standard
                deviation of the metallicity for the disk, stream, and halo11
                fields are
                consistent with the expected accuracy of the method
                ($\sigma_{[Fe/H]} \sim$ 0.31 and 0.45, respectively).  No
                errors are given for metallicities using the c-type variables,
                as no error estimates were estimated or provided for the
                relation in the original source.}
      \label{calctable}
\end{table*}

   For ab-stars with short periods, this relation can produce supersolar 
metallicities.  We note that it is for this reason that we exclude star 7085 in 
the halo35b field throughout this analysis (see Section \ref{period}) as its  
short period would produce such an outlying metallicity.  While 
such a value may be valid, S06 note that their metallicity relation was not 
derived for such high metallicities.  While this star is still used when 
normalizing the sparse distribution of this field, we will disregard it in our 
subsequent analysis of this field, as it would highly skew any statistics, due 
to the already small sample of this field.  In particular, we note that the 
average [Fe/H]$_{RR{ab}}$ value for halo35b in Table \ref{calctable} excludes 
this outlier, and we also exclude it in the distance calculation for this field.

   It is important to understand the limitations of Equation 1 when 
determining [Fe/H] values for individual RR\textit{ab} stars.  First, as noted 
above, the rms dispersion in the [Fe/H] values given by this equation is 
very large, and thus any individual determination will be quite 
uncertain.  More importantly, Equation 1 is inconsistent with the 
fundamental pulsation equation, since it contains no information about 
the location of an RRab star within the instability strip.  According to 
van Albada \& Baker (1971) the pulsation period of an RR$_{ab}$ star depends 
on the mass $M$, the luminosity $L$ and the effective temperature $T_{eff}$ as 
follows:

\begin{equation}
   \log{P_{ab}}  =  11.497  -  0.68 \log{M} +  0.84 \log{L} -  3.48 \log{T_{eff}} 
\end{equation}

\noindent
where $M$ and $L$ are in solar units.  Thus an RR\textit{ab} star located near 
the fundamental red edge of the instability strip would have a longer period 
and thus smaller [Fe/H] value according to Equation 1 than an RR$_{ab}$  
star with the same mass and luminosity located near the fundamental blue 
edge.  This is not a small effect.  Consider, for example, the OoI 
cluster M3 for which there is no observational evidence for a 
significant variation in [Fe/H].  HB simulations show that the bulk of 
RR\textit{ab} stars in M3 have very similar masses and luminosities.  Thus the 
variation in $\log{P_{ab}}$ across the instability strip in M3 comes primarily 
from the variation in $T_{eff}$.  Canonical HB tracks show that the 
RR\textit{ab} stars in M3 are evolving blueward along blue loops as they cross 
the instability strip.  Equation 1 would therefore imply that the [Fe/H] 
values for such stars should increase during their evolution even when 
their masses and luminosities are constant.  The size of this effect can 
be estimated from the pulsation periods obtained by Corwin \& Carney 
(2001) in their study of the M3 RR Lyrae stars.  Their data show that 
most RR\textit{ab} stars in M3 have periods between $\log{P_{ab}}$ = -0.15 
and -0.35, implying a difference of 1.5 dex in [Fe/H] between the reddest and 
bluest RR\textit{ab} stars in contradiction to the observed abundances.  It is 
not surprising therefore that the [Fe/H] values obtained from Equation 1 
show such a large dispersion among the field RR\textit{ab} stars.

   We do however recognize that this relation is usually only meant to be used 
when the amplitudes of RR Lyraes are very uncertain or unavailable.  When this 
relation is the only viable option, however, it is important that authors 
proceed with great caution, given these great uncertainties.

   When calculating the distance modulus, $(m-M)_{0}$, for each field, we must 
first determine the average apparent magnitude of each star in the $V$ band. 
The transformation of the F606W to the $V$ band was done as we described in 
Section \ref{ootype}.  

   Carretta et al. (2000) derive a relationship between an RR Lyrae star's 
metallicity and its absolute magnitude.  Once we know the metallicities of the 
RR Lyrae stars, we can determine its absolute magnitude using their relation:

\begin{equation}
   M_{V} = (0.18 \pm 0.09) * ([Fe/H] + 1.5) + (0.57 \pm 0.07).
\end{equation}
   \label{disteq}

\noindent
We use this relation and the average metallicity determinations for the 
RR\textit{ab} stars to calculate the average absolute magnitude of the RR
Lyraes in each field.  We then subtract it from the average unreddened $V$ 
magnitude of each respective field to determine the unreddened distance 
modulus.  For the disk, stream, and halo11 fields the reddenings used were those 
calculated as outlined in Section \ref{red}.  Reddening for the halo21 and 
halo35b was assumed to be $E(B-V)$ = 0.08 $\pm$ 0.03 because of the low number 
of RR Lyrae in these two fields.  (This is a valid assumption, given the low 
amounts of dust in the halo.)  We list the result of this in Table 
\ref{distcalc}.


\begin{table*}
  \begin{center}
  \begin{tabular}{cccccc}

    \hline

 $(m-M)_{0}$   & disk & stream & halo11 & halo21 & halo35b  \\
    \hline


 S06      & 24.50 $\pm$ 0.22 & 24.56 $\pm$ 0.22 & 24.49 $\pm$ 0.23  & 24.37 $\pm$ 0.29 & 24.35 $\pm$ 0.24 \\
          & \multicolumn{1}{r}{(794)} & \multicolumn{1}{r}{(817)} & \multicolumn{1}{r}{(791)} & \multicolumn{1}{r}{(748)} & \multicolumn{1}{r}{(741)} \\
 A00      & 24.50 $\pm$ 0.22 & 24.60 $\pm$ 0.23 & 24.51 $\pm$ 0.23 & 24.34 $\pm$ 0.29 & 24.35 $\pm$ 0.23 \\
         & \multicolumn{1}{r}{(794)} & \multicolumn{1}{r}{(831)} & \multicolumn{1}{r}{(798)} & \multicolumn{1}{r}{(738)} & \multicolumn{1}{r}{(741)} \\
 B07(1.5) & 24.34 $\pm$ 0.12 & 24.47 $\pm$ 0.20 & 24.40 $\pm$ 0.08 & \--- & \--- \\
          & \multicolumn{1}{r}{(738)} & \multicolumn{1}{r}{(783)} & \multicolumn{1}{r}{(759)} & \---                      & \---                      \\
 B07(2.0) & 24.51 $\pm$ 0.12 & 24.62 $\pm$ 0.18 & 24.55 $\pm$ 0.07 & \--- & \--- \\
          & \multicolumn{1}{r}{(798)} & \multicolumn{1}{r}{(839)} & \multicolumn{1}{r}{(813)} & \---                      & \---                      \\
 RR$_{c}$ & 24.61 $\pm$ 0.12 & 24.43 $\pm$ 0.12 & 24.47 $\pm$ 0.12 & \--- & 24.41 $\pm$ 0.17 \\
          & \multicolumn{1}{r}{(836)} & \multicolumn{1}{r}{(769)} & \multicolumn{1}{r}{(783)} & \---                      & \multicolumn{1}{r}{(762)} \\
 Weighted & 24.49 $\pm$ 0.17 & 24.52 $\pm$ 0.19 & 24.48 $\pm$ 0.15 & 24.36 $\pm$ 0.40 & 24.38 $\pm$ 0.27 \\
 Average  & \multicolumn{1}{r}{(791)} & \multicolumn{1}{r}{(802)} & \multicolumn{1}{r}{(787)} & \multicolumn{1}{r}{(745)} & \multicolumn{1}{r}{(752)} \\


   \hline
   \end{tabular}
 \end{center}
   \caption{Summary of the calculated distances of the RR Lyraes in our
                M31 fields.  The method used is indicated in the first column.
                The B07(1.5) and B07(2.0) methods indicate the results using
                $\alpha$ = 1.5 and 2.0, respectively.  Errors of the distance
                calculated from the c-type stars (last row) are much smaller
                than the other methods because they do not include error in
                metallicity.  The numbers in the parenthesis below each
                distance modulus value is that distance modulus value
                converted to Mpc.}

      \label{distcalc}
\end{table*}

   The error of the absolute magnitude for each star is calculated by 
propagating the errors of Equation 3, and assuming an error on metallicity 
(standard deviation) for 
each field as listed in Table \ref{calctable}.  Errors in the average 
unreddened apparent magnitude for each field are calculated as the standard 
error of the mean of the sample.  The final quoted errors of Table 
\ref{distcalc} are done by adding in quadrature the errors on apparent 
magnitude, absolute magnitude, and (for the halo21 and halo35b) absorption.  

   Our results are in agreement within the error bars with previous distance 
determinations of M31 using RR Lyraes.  For example, Sarajedini et al. (2009) 
find a distance of $(m-M)_{0}$ = 24.46 $\pm$ 0.11.  (We note that our errors 
are larger because we propagated our metallicity error to the end.)  Our 
results also agree with the Cepheid distance to M31, $(m-M)_{0}$ = 24.44 $\pm$ 
0.1 (Freedman \& Madore 1990).

\subsection{The Period-Amplitude-Metallicity Relation}
   \label{alcock}

   Among the most frequently used relations to determine the metallicity of 
RR\textit{ab} stars was found by Alcock et al. (2000, hereafter A00).  They 
found a  period-amplitude-metallicity relation for ab-type stars in the form


\begin{equation}
   [Fe/H]_{RR_{ab}} = -8.85[\log{P_{ab}} + 0.15A(V)] - 2.60,
\end{equation}
   \label{abfeheq}

\noindent
where $A(V)$ is the amplitude in the $V$ band and $P_{ab}$ is the period in
days.  This relation has been used in many previous studies, including Paper 
I and Sarajedini et al. (2009), and is similar to the relation found by 
Sandage (2004).  When applying this relation to our data, we 
calculated the $V$ band amplitude as described in Section \ref{ootype}.

   We plot the normalized MDF for each field in the middle row of panels in 
Figure \ref{fehdist}.  The average of the MDF for each field is given in Table 
\ref{calctable}.  Again, two errors are quoted for each field, the standard 
deviation and the standard error of the mean (in parenthesis).  And as before, 
these values for the halo21 and halo35b fields use the formulae appropriate 
for small statistical samples. 

   A00 note that the accuracy of this technique is $\sigma_{[Fe/H]} \sim$ 
0.31.  Standard deviations of the MDF for the disk, stream, and halo11 fields 
(i.e., the fields with a substantial enough RR Lyrae population to produce a 
well-populated MDF) are consistent with the accuracy of the technique.

   Sarajedini et al. (2009) use the A00 relation to determine the metallicities 
of a sample of RR Lyrae stars in the spheroid of M31 and compare the resulting 
MDF to the MDF found using the S06 relation (Equation 1).  They find the 
results to be similar.  We also find these two distributions to be similar 
when applied to our sample. 
The average metallicity values from the A00 method agree within the error bars 
to the S06 metallicities.  As before, we excluded star 7085 in the halo35b 
field from the analysis (see Section \ref{ata}).

   At first glance the A00 relation for the RR\textit{ab} metallicity would 
appear to be an improvement over the S06 relation, because it includes a 
term involving the amplitude $A(V)$ which provides a measure of a star's 
location within the instability strip.  Unfortunately, this is not the 
case.  As emphasized by Bono et al. (2007), the form of 
any period-amplitude-metallicity relation depends strongly on the choice 
of calibrating clusters (see their Figure 5).  The A00 relation was 
calibrated on the clusters M15, M3 and M5.  However, the use of other 
clusters can yield a quite different calibration and hence quite 
different values of [Fe/H].  Essentially the A00 relation is an attempt 
to determine the metallicity from the size of the period shift in the 
Bailey diagram of the RR\textit{ab} stars.  However, Figure 5 of Bono et al. 
(2007) shows, for example, that the period shift is negligible among the OoI 
globular clusters over a wide range in [Fe/H] from -1.8 to -1.1.

   Once metallicities are found, we calculate absolute magnitudes and distance 
moduli for each field, using the same method as described in the previous 
section.  Results are listed in Table 16.

\subsection{The Bono Method}
   \label{bono}

   The relations we have discussed thus far have been frequently used by RR 
Lyrae researchers in numerous studies, and are therefore useful for putting 
our results into a similar context for comparison.  However, the physical 
reliability of these relations has been recently 
questioned by a number of studies in the literature.

   A recent paper by Catelan (2009) 
shows that there is no clear trend in the mean RR$\textit{ab}$ period with 
metallicity for the OoI clusters, contrary to what the A00 and S06 relations 
indicate (i.e., that an increase in period leads to a decrease in metallicity, 
at constant $A(V)$).  The errors of derived [Fe/H] values from both the A00 
and S06 relations are high, which lead to high errors in distance when properly 
propagated through the relations between metallicity and absolute magnitude.  
(Often RR Lyrae researchers drop the absolute error bars on 
metallicity when calculating errors of absolute magnitude, and hence distance 
modulus, reporting distance errors that are much smaller than one would 
expect, given the large error in metallicity.)  

   Bono et al. (2007, hereafter B07) addressed some of the uncertainties in the 
RR Lyrae relationships by 
deriving equations relating the observed quantities of these pulsators (e.g., 
RR\textit{ab} period and $A(V)$) to actual physical properties of the star 
(e.g., luminosity and mass).  They did this using nonlinear hydrodynamical 
models for the RR\textit{ab} pulsation.  These equations depend modestly on the 
assumed mixing length ratio $\alpha$ in the models, so they considered two 
values, $\alpha$ = 1.5 and 2.0.  Their analysis yielded relations between 
M$_{V}$ and the RR\textit{ab} periods and $A(V)$.  Given the values of the 
M$_{V}$, one can then derive the metallicity by adopting an appropriate 
metallicity-M$_{V}$ relation.  This method is therefore the reverse of the 
methods discussed above: rather than first calculating metallicities and then 
absolute magnitudes (and hence distances), this technique first calculates the 
absolute magnitude and then the metallicity.

   In their study, B07 define a parameter $k_{puls}$ for both values of 
$\alpha$ as

\begin{equation}
  k(1.5)_{puls} = 0.136 - \log{P_{ab}} - 0.189A(V)
\end{equation}
\begin{equation}
  k(2.0)_{puls} = 0.027 - \log{P_{ab}} - 0.142A(V) 
\end{equation}

\noindent
Additionally, they find that the average absolute 
magnitude, $<M_{V}>$, of a population's RR Lyraes stars is related to the 
average of the $k_{puls}$ parameter ($<k_{puls}>$), in the following relations 
(their Equations 5 and 6), again, listed for both mixing length values:

\begin{equation}
  <M_{V}^{k(1.5)}> = 0.12(\pm 0.10) + 2.65(\pm 0.07)<k(1.5)_{puls}>
\end{equation}
\begin{equation}
  <M_{V}^{k(2.0)}> = 0.14(\pm 0.10) + 2.67(\pm 0.07)<k(2.0)_{puls}>.
\end{equation}

\noindent
For metallicities greater than $\--$1.0, they give the following relations (their 
Equations 8 and 9):

\begin{equation}
 M_{V}^{k(1.5)} = 0.56 - 0.49A(V) - 2.60 \log{P_{ab}} + 0.05[Fe/H] 
\end{equation}
\begin{equation}
  M_{V}^{k(1.5)} = 0.64 - 0.49A(V) - 2.60 \log{P_{ab}} + 0.20[Fe/H]
\end{equation}

\noindent
 for -1.0 $\leq$ [Fe/H] $\leq$ -0.5 and -0.5 $\leq$ [Fe/H] $\leq$ 0.0, 
respectively.

   The above Equations 7 and 8 can be used to find the distance modulus of the 
metal poor stars (i.e., [Fe/H] $\leq \--$1.0, where the relations are valid) 
in the population of interest.  B07 did this for the globular cluster $\omega$ 
Cen, for both possible values of the mixing length.  They plotted the 
unreddened visual magnitudes of the cluster RR\textit{ab} stars against the 
$k_{puls}$ parameter for each star (see their Figure 17).  These data are then 
fit with a straight line with the slope of the above Equations 7 and 8.   The 
difference between this line and the line defined by these equations, i.e., 
the shift in magnitude needed to match the data, is the distance modulus.  The 
distance values found by B07 for $\omega$ Cen using the different $\alpha$ 
values bracket the accepted distance value, indicating that the proper value 
for $\alpha$ is likely near 1.7.  Unfortunately, however, no equations were 
provided for this best value.

\begin{figure}
   \epsscale{1.20}
        \plotone{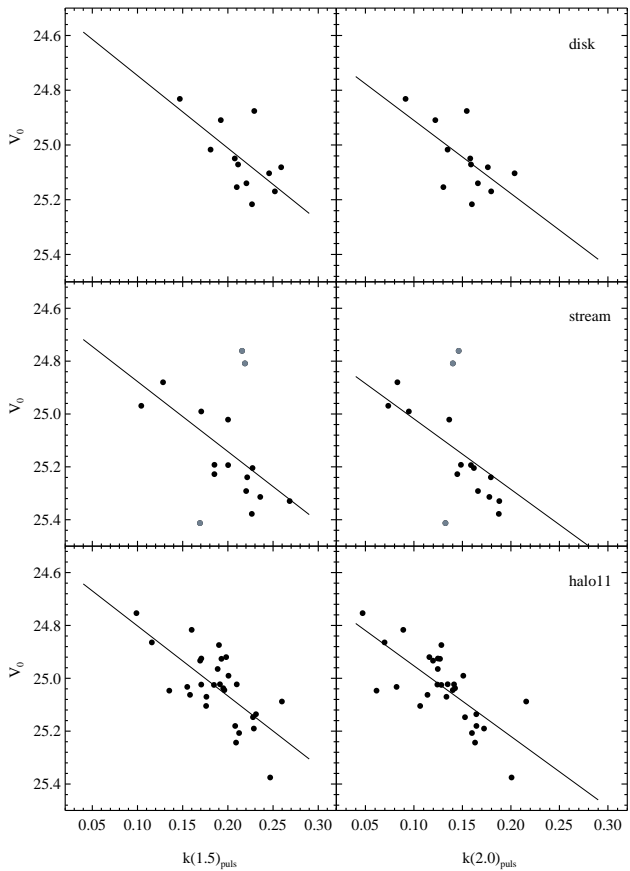}
        \caption{Unreddened visual magnitude $V_{0}$ of RR$_{ab}$ in the disk,
                stream, and halo11 fields, plotted versus $k$(1.5)$_{puls}$
                (left panels) and $k$(2.0)$_{puls}$ (right panels).  The solid
                lines are Equations 6 and 7 (in the respective panels) and
                account for the intrinsic distance modulus.  The distance
                modulus found from the best fit is listed in Table
                \ref{distcalc}.  See text for discussion.}
   \label{kpuls}
\end{figure}

   We have done this same analysis for the ab-type stars in our disk, stream, 
and halo11 fields, where the population sample is big enough to get meaningful 
results.\footnote{As we will show in Section \ref{layden}, it is possible to apply 
this method to individual stars; however, deriving a distance modulus from the 
method of fitting a straight line to $V_{0}$ vs. $k_{puls}$ plot, requires a 
large enough sample, as we have here.}  In Figure \ref{kpuls} we have plotted the unreddened 
magnitude values for the RR\textit{ab} stars (using the individual reddening 
values found in Section \ref{red}) and then calculated the corresponding 
$k_{puls}$ values.  The panels on the left are for $\alpha$=1.5, while those 
on the right are for $\alpha$=2.0.  The solid line is a best fit (using a 
standard linear least squares fitting technique) of a straight line with the 
slope of Equations 7 and 8 (on the respective plots).  The best fit distance 
modulus in each case is listed in Table \ref{distcalc}, for both values of 
$\alpha$.  The accompanying error is the error of the fit.  Similar to the B07 
results, our results likely bracket the true value which probably lies within 
the error bars of both.


   Because this method of finding the distance is independent of a calculated 
metallicity, we note that the errors are lower than the previous two methods 
(although they all agree within the errors).  
Not only that, it relies on a method that is much more physical, and therefore 
likely more reliable.  We note the extreme outlying points in the stream field 
(the gray points) leading to higher distance errors in this field compared to 
the disk or halo11 field.  The two high outlying points are stars 561 and 2933, 
while the low outlying point is star 2577.  The fits of the light curves of 561 
and 2933 (see Figure \ref{strmLC}) are poorer than for the other stars, 
possibly explaining why these stars are outliers.  Despite this, we retain 
these stars in our analysis.

   In addition to the distance calculations, B07 derive a relation between 
metallicity and absolute magnitude, valid over the 
metallicity range of [Fe/H] = $\--$2.5 to $\sim$0.  The relation is

\begin{equation}
   M_{V}^{k(1.5)} = 1.19(\pm 0.10) + 0.50[Fe/H] + 0.09[Fe/H]^{2}.
\end{equation}

   Using this relation, we calculated individual [Fe/H] values from the values 
of $M_{V}^{k(1.5)}$ given by Equation 7 for the RR Lyrae stars in each of these 
three fields.  The MDF constructed in this manner is shown as the bottom row 
of panels in Figure \ref{fehdist}.  We have listed the average, standard 
deviation, and standard error of the mean of these MDFs in Table 
\ref{calctable}, as before.

   We note that errors on metallicity for our fields using the B07 method are 
mostly comparable to those of the A00 and S06 methods.  However, one major 
advantage of the B07 method is that these metallicity errors do not propagate 
into distance errors as they do with the other methods.

   Our results for distance using the method of B07 are consistent within the 
errors with those found in the previous section, which is encouraging.  The 
metallicity values are overall slightly more metal rich (compare values in 
Table \ref{calctable}).  
However, because the B07 results were derived using a more physically 
meaningful, consistent way, we believe they are a better representation of the 
actual physical values of this sample of RR Lyraes.  Relative metallicities 
remain nearly the same when compared to the other two methods. 

   B07 also found that the distribution of the $k_{puls}$ parameter can 
potentially be a useful indicator of the Oosterhoff type.  They determined that 
the location of the peak in the distribution of $k_{puls}$ is dependent on its 
Oosterhoff type.  Also, they note that the distribution of OoII clusters is 
slightly different for clusters that have a horizontal branch (HB) bluer than 
+0.8\footnote{This parameter is defined as $(B - R)/(B+V+R)$ between 
the numbers of HB stars to the blue (B), within (V), and to the red (R) of the 
RR Lyrae instability strip (Lee 1990).} when compared 
to those whose HBs do not extend that far.  

\begin{figure}
   \epsscale{1.20}
        \plotone{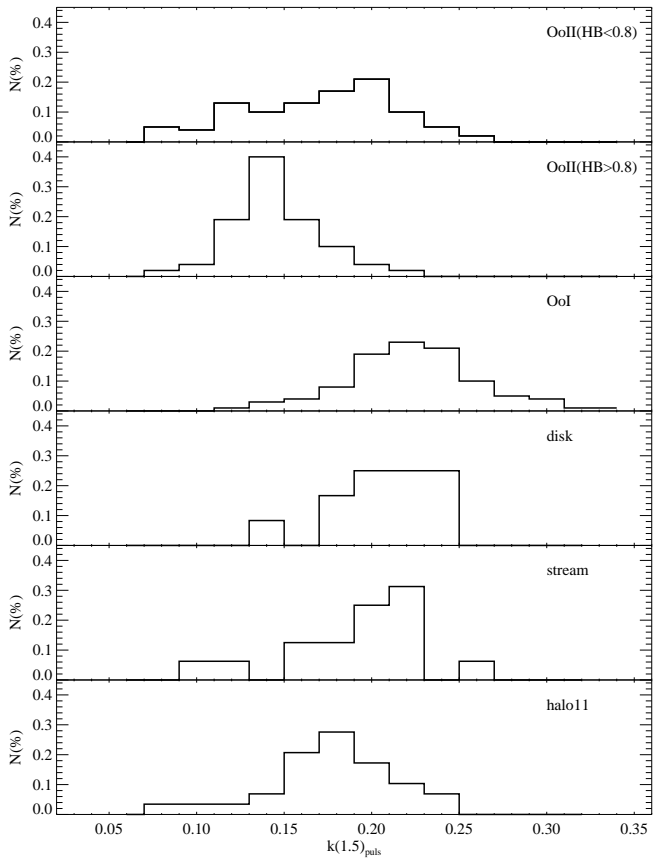}
        \caption{The distribution of the $k(1.5)_{puls}$ parameter for MW GCs
        of different Oosterhoff type (as labeled), as well as for the disk,
        stream, and halo11 fields.  As is evident from this plot, we confirm
        that the stream and the disk are likely OoI type, while the halo11
        field is an intermediate population, or possibly a mixed population of
        OoI and OoII.}
   \label{kpulshist}
\end{figure}

   We plot these distributions in Figure \ref{kpulshist}.  The top three 
panels of this plot are taken from B07 (see their Figure 8), and show the 
distribution of $k(1.5)_{puls}$ for OoI and OoII clusters, as labeled.  The 
bottom three panels are these distributions for our disk, stream, and halo11 
fields.  As can be seen from this plot, the stream and the disk are indeed 
more similar to the OoI type (given the peak and shape of their distributions) 
while the halo11 field is an intermediate population.  We note however, that 
it is possible that the halo11 field is a mixed population, with some 
combination of OoI and OoII.  This scenario would be consistent with the 
findings of Sarajedini et al. (2009) who find fields in the halo of M31 at 4 
and 6 kpc to be OoI.


\subsection{Comparison with the Layden Data}
  \label{layden}

   In order to test the previous three methods, we have compared the 
metallicities given by these methods with those determined spectroscopically 
for 78 field ab-type RR Lyrae stars from an unpublished dataset from A.C. 
Layden (private communication).  This will allow us to use the measured 
periods and amplitudes to determine how well the different methods reproduce 
the actual metallicities of these stars.

\begin{figure*}
   \epsscale{1.00}
        \plotone{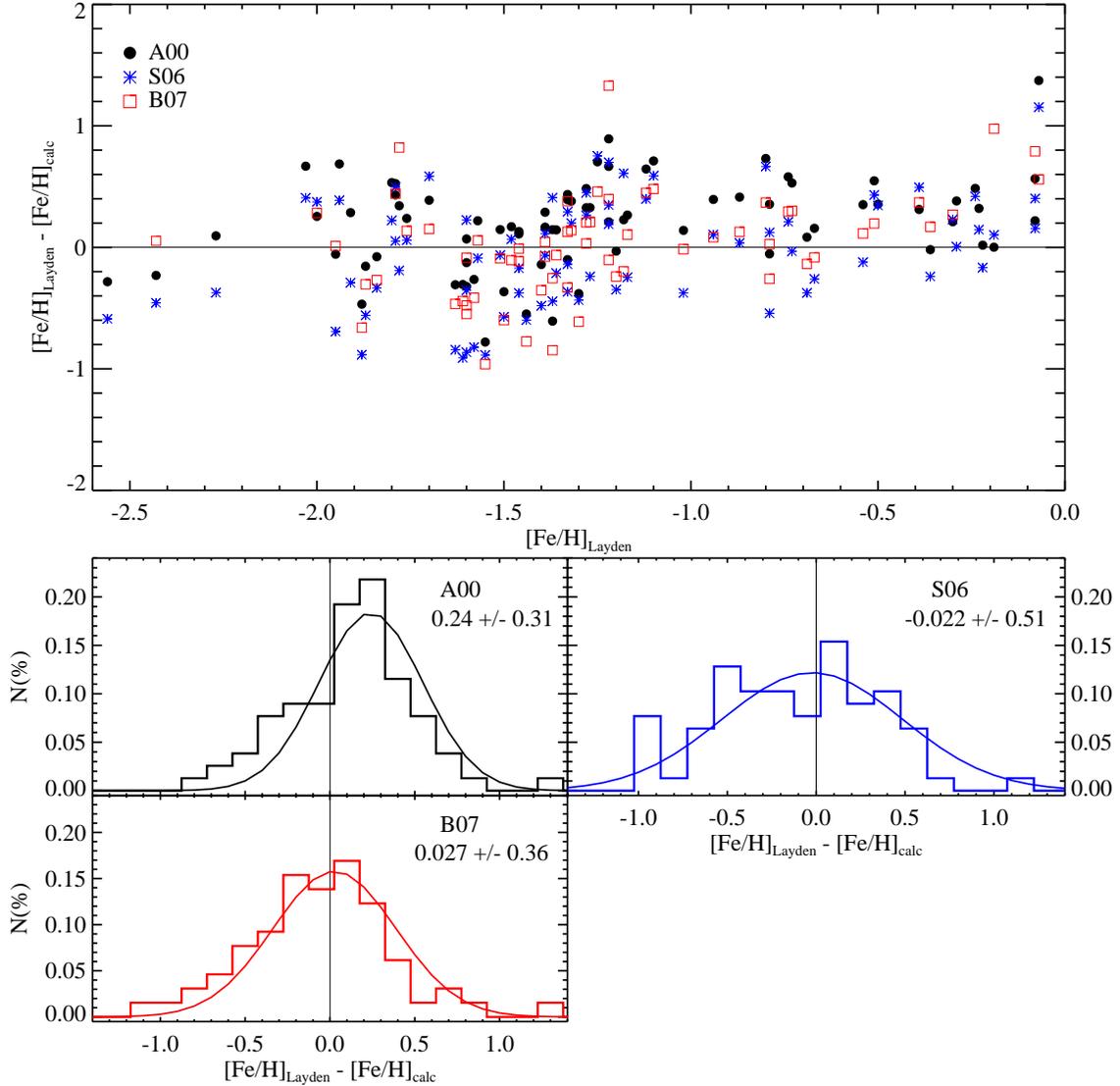}
        \caption{A comparison of the three methods we have used for
        calculating RR Lyrae metallicities.  The top plot is the
        difference in the metallicity calculated minus the spectroscopic value
        vs. the spectroscopic metallicity, with the colors and symbols as
        indicated.  The horizontal solid line is at a difference equal to 0.
        The panels in the bottom plot are histograms of the above differences,
        with a best fit Gaussian to each distribution overplotted.  The
        numbers on the plot indicate the average and standard deviation of
        the Gaussian, and aid in quantifying how well each method can
        reproduce the ``true" metallicity.}
   \label{layden_plot}
\end{figure*}

   We display the results of this in Figure \ref{layden_plot}.  The top plot 
is the difference in metallicity calculated by each method minus Layden's 
spectroscopic value as a function of Layden's metallicities, with the color and 
symbols as indicated.  (We remind the reader that the B07 metallicities 
are calculated using only $\alpha$ = 1.5, as they do not provide a metallicity 
relation for $\alpha$ = 2.0.  Additionally, we note that we used the proper 
$M_{V}$ relation, given the metallicity of each star, see Equations 7, 9, and 
10, when solving for the metallicity, Equation 11.)  The horizontal solid line 
is at a difference equal to 0.


   The three panels in the bottom plot of this same figure are histograms of 
the above differences, giving us an idea of how well each method reproduces 
the spectroscopic values.  Again, the solid vertical line is at difference of 
0.  We have fit a Gaussian distribution to each histogram and overplot it to 
help quantify the method's ability to reproduce the metallicities and the 
error.   The numbers on the plot indicate the average and standard deviation 
of the Gaussian.

   The average difference of each of the three methods comes out to nearly 
zero, which is encouraging.  The S06 (period-metallicity) method does the 
poorest job of the three at reproducing the spectroscopic metallicity values, 
and it has the largest spread ($\sigma \sim$ 0.5).  However, this isn't 
surprising because it has the least information going into the calculation 
(i.e., it uses periods alone, not amplitudes).  The A00 
(period-amplitude-metallicity) method does a better job with a tighter 
distribution, although it is systematically off by nearly 0.25 dex.  This is 
likely because of the choice of calibrating GCs in determining the 
zero point and slope in the relation (see 
Equation 4).  The errors we find on these two methods are consistent with the 
expected uncertainty for an individual star ($\sim$ 0.31 and 0.45 for A00 and 
S06, respectively.)

   The B07 method does the best job of reproducing the spectroscopic 
metallicities, closely centered around zero difference, although the A00 and 
S04 methods both agree within the errors.  The spread in these distributions is 
roughly consistent with what we see in our M31 data (see Table 
\ref{calctable}).  However, because metallicity errors of the B07 method do not 
propagate through to the absolute magnitude (and hence distance) calculations, 
and the derivation is more physical, B07 is our preferred method.

\subsection{The c-type Stars}
  \label{ctype}

   The RR\textit{c} stars also exhibit a relation between period and 
metallicity.  However, unlike the ab-type stars, this relation relies on the 
average period of the c-type variables in a given population, rather than 
individual stellar properties.  This is calculated using the relation (from 
Sandage 1993)


\begin{equation}
   [Fe/H]_{RR_{c}} = (-\log{<P_{c}>} - 0.670) / 0.119
\end{equation}
   \label{cfeheq}

\noindent
where $<P_{c}>$ is the average period of the c-type RR Lyrae stars.  Because 
this method calculates metallicity from the mean properties of a population, a 
single number for each field is determined, rather than a distribution like 
the ab-type stars.  We note the value calculated by this method in Table 
\ref{calctable}.  We also note that we have not quoted errors for the 
metallicity calculated from the RR\textit{c} stars, as no error estimates were 
provided for the above equation.  Given this metallicity, we then calculated 
the distance to each field using Equation 3, in the same manner as before.  
These values are listed in Table \ref{distcalc}.

\vspace{5 mm}

   Regardless of the method used, we desire to compare the metallicities 
calculated here with those calculated by detailed studies of the CMDs of these 
fields (Brown et al. 2006), as well as spectroscopic work (e.g., Guhathakurta 
et al. 2006).  These studies found the populations to be, on 
average, more metal rich than calculated here by about 0.5 to 1.0 dex.  The 
reason for this is that the RR Lyrae stars pre-select towards the metal poor 
tail of the respective population distributions.  However, we note that the 
relative metallicities (at least those calculated by methods that use the 
RR\textit{ab} stars) are similar: e.g., the disk is more metal rich than the 
halo and the stream, and the stream is more metal rich than the halo.  

\section{Summary and Conclusions}
  \label{conclusion}

   We have presented a complete survey of the RR Lyrae stars in six ultra deep 
fields in various parts of M31: namely, the disk, the giant stellar stream, 
and halo fields of varying distance from the galactic center.  We find RR 
Lyraes in five of the six fields.  The halo11 field is of intermediate (or 
mixed) Oosterhoff type, the disk and the stream are of OoI type; the halo21 and 
halo35b fields seem to have a tendency toward OoI type, but are too sparse to 
meaningfully classify as one type or the other.  We determine the average 
reddening of the disk, stream, and halo11 fields and find that the disk is 
more reddened than the literature value to M31, while the halo11 is roughly 
the same.

   We have calculated metallicities and distances for each field of our 
sample, using a variety of methods including the period-metallicity relation 
of Sarajedini et al. (2006), the period-amplitude-metallicity of Alcock et al. 
(2000), and the method of Bono et al. (2007) that is derived from more 
physically realistic models (using two values for the mixing length), rather 
than simple empirical relations.  These three methods are roughly consistent, 
although the latter method has smaller errors than the former on distance 
though errors are similar for metallicity.

    To compare these three methods for calculating metallicities, we have 
applied each of these methods to data of MW field RR Lyraes.  We found that 
although the three methods tend to overlap and be consistent within the 
errors, the B07 method is our preferred method.  It not only provides closest 
match and is based on more physical derivations, but importantly, high 
metallicity errors do not propagate through to distance calculations, as with 
the more traditional methods.

\begin{acknowledgements}

 \begin{center}
\textbf{Acknowledgments}
  \end{center}

   We wish to thank A.C. Layden for kindly providing us with his data for the 
field RR Lyrae stars.  We would also like to thank Ata Sarajedini for helpful 
discussions in the later stages of this paper.  We acknowledge support for 
Programs GO-9453, GO-10265, and GO-10816 was provided by NASA through a grant 
from the Space Telescope Science Institute, which is operated by the 
Association of Universities for Research in Astronomy, Incorporated, under 
NASA contract NAS5-26555.  We are also grateful for the useful comments of an 
anonymous referee, which helped to clarify this paper.

\end{acknowledgements}

\end{document}